\documentclass[aps,pra,epsfigure,twocolumn,longbibliography]{revtex4-1}
\usepackage{dcolumn}    
\usepackage{bm} 
\usepackage{graphicx}
\usepackage{amsmath}
\usepackage{mathtools}
\usepackage{latexsym}
\usepackage{amsfonts}   
\usepackage{amssymb}
\usepackage{array}      
\usepackage{epsfig}
\usepackage{txfonts}
\usepackage{color}
\usepackage[colorlinks=true,linkcolor=blue,urlcolor=blue,citecolor=blue,pdfusetitle]{hyperref}
\usepackage{hyperref}
\usepackage[normalem]{ulem}

\newcommand{\ket}[1]{\left\vert#1\right\rangle}
\newcommand{\bra}[1]{\left\langle#1\right\vert}
\newcommand{\nbar}{\bar{n}}

\catcode`\>=\active \def>{
\fontencoding{T1}\selectfont\symbol{62}\fontencoding{\encodingdefault}}
\newcommand{\tmop}[1]{\ensuremath{\operatorname{#1}}}
\newcommand{\tmtextit}[1]{{\itshape{#1}}}

\begin{document}
\title{Precursors of non-Markovianity}
       
\author{Steve Campbell,$^{1,2}$ Maria Popovic,$^2$ Dario Tamascelli,$^{3,4}$ and Bassano Vacchini,$^{3,1}$}
\affiliation{
$^1$Istituto Nazionale di Fisica Nucleare, Sezione di Milano, via Celoria 16, 20133 Milan, Italy\\
$^2$School of Physics, Trinity College Dublin, Dublin 2, Ireland\\
\mbox{$^3$Dipartimento di Fisica ``Aldo Pontremoli", Universit{\`a} degli Studi di Milano, via Celoria 16, 20133 Milan, Italy}\\
$^4$Institute of Theoretical Physics, Universit\"at Ulm, Albert-Einstein-Allee 11, D-89069 Ulm, Germany}

\begin{abstract}
Using the paradigm of information backflow to characterize a non-Markovian evolution, we introduce so-called precursors of non-Markovianity, i.e. necessary properties that the system and environment state must exhibit at earlier times in order for an ensuing dynamics to be non-Markovian. In particular, we consider a quantitative framework to assess the role that established system-environment correlations together with changes in environmental states play in an emerging non-Markovian dynamics. By defining the relevant contributions in terms of the Bures distance, which is conveniently expressed by means of the quantum state fidelity, these quantities are well defined and easily applicable to a wide range of physical settings. We exemplify this by studying our precursors of non-Markovianity in discrete and continuous variable non-Markovian collision models.
\end{abstract}
\date{\today}
\maketitle

\section{Introduction}
Open quantum systems provide the framework for describing how a system of interest interacts with its surroundings. The ubiquity of this paradigm has led to a wide variety of techniques to model, characterize, and exploit the manner in which an environment affects a system's evolution~\cite{Alicki1987,Weiss1993,Breuer2002,Rivas2012}. In the context of open quantum systems, typically the system of interest is much smaller than its environment and in this case, if the coupling is weak, the dynamics of the system can be well captured by a memoryless or Markovian evolution.

The failure of such an approximation in the presence of strong coupling, structured reservoirs, or non-negligible system effects on the environment, naturally leads us to explore dynamics with some form of memory. Dynamics of this sort are typically referred to by the catch-all term non-Markovianity and have been the subject of intense activity, evidenced by the development of a range of techniques to simulate and characterize a wide range of non-Markovian dynamics, see e.g. the recent reviews~\cite{Breuer2012a,Breuer2016a,Rivas2014a,Devega2017a,Li2018a}. Indeed beyond being a topic of interest in itself, recent work has shown that non-Markovianity of the dynamics can provide an enhancement in a diverse array of settings and tasks, including quantum-metrology, quantum memories, information processing, and thermodynamic cycles~\cite{Plenio1, Plenio2, DeffnerPRL2015, PezzuttoQST, BarisArXiv, ModiJPhysA, NJPQI}.

Despite the clear relevance of the field and its success in showing potential applications, understanding the fundamental mechanisms and features that give rise to quantum non-Markovianity remains a difficult task. Indeed the very foundations of the theory are still the object of intense investigations~\cite{Pollock2018a, Modiarxiv}, the reason for which is due to the fact that different approaches to defining non-Markovianity have been devised, whose relationship is currently under study \cite{Chruscinski2011a,Wissmann2015a,Breuer2018a,Chruscinski2018a}.

In this work we attempt to underpin some fingerprints of a non-Markovian evolution by considering precursors of non-Markovianity, i.e. features of the system and environment state at some time, $s$, that serve as necessary (but not sufficient) indicators that the dynamics can show a non-Markovian behavior at a later time $t\!>\!s$.  These quantities can be introduced in the framework of non-Markovianity described as a backflow of information from the environment to the system~\cite{BreuerPRL2009, Breuer2016a}. In this picture, bounds can be introduced that explicitly relate the behavior of a particular indicator of non-Markovianity with the features of the joint system-environment state~\cite{Laine2010b,Amato2018a}. Precursors of non-Markovianity provide necessary conditions for tracing revivals in a suitable distinguishability quantifier back to a combination of established system-environment correlations and changes in the environmental state. A significant limitation of the approach is that it calls for evaluation of features of both the system and the {\it total} environment, which is generally too demanding a task, and as such investigations along these lines have rarely been considered~\cite{Mazzola2012a,Rodriguez2012a,Smirne2013b}. However, it has been shown that when the environment can be described as a collection of individual subsystems, often only a small subset of the total environmental degrees of freedom plays a significant role in driving the evolution~\cite{NazirPRA2014, CampbellPRA2018}.  By relying on the Bures distance and considering collision models of open quantum system dynamics~\cite{rau,brun,ScaraniPRL2002, BuzekPRA2005}, where the environment is modelled by an ensemble of individual ancillae with which the system sequentially interacts, we are able to introduce and study a hierarchy of precursors of non-Markovianity for systems in both discrete and continuous variable settings. Our work paves the way for considering the relevance of such precursors in a variety of settings which allow for the systematic introduction of environmental degrees of freedom, e.g. within the framework of reaction coordinate models or chain mappings \cite{Hughes2009a, Chin2010a, Martinazzo2011a, Woods2014a, Strasberg2016a, Tama1, Tama2}.

The remainder of the paper is organized as follows. In Sec.~\ref{precursors} we define the precursors of non-Markovianity, and in particular the bound, Eq.~\eqref{eq:ineqB}, that is central to our analysis. In Sec.~\ref{collisionmodel} we apply our framework to both discrete and continuous variable collision models, and qualitatively compare and contrast the emergence of non-Markovianity in the disparate dimensional systems. Finally, in Sec.~\ref{conclusions} we draw our conclusions.

\section{Precursors of non-Markovianity}
\label{precursors}
In a seminal work, the non-Markovianity of a reduced quantum dynamics has been introduced as an indicator of information backflow from the environment to the system~\cite{BreuerPRL2009}. In order to detect and estimate this backflow a distance on the state space, namely the so-called trace distance, has been considered. Apart from a normalization factor, this distance, given for any pair of states $\rho$ and $\sigma$ by
\begin{eqnarray}
D ( \rho , \sigma ) & = & \frac{1}{2} \| \rho - \sigma \|_{1} , \nonumber
\end{eqnarray}
corresponds to the natural topology on the state space and has two basic features, which makes it suitable as an estimator of information backflow: \tmtextit{(i)} it is directly related to a notion of distinguishability among states; \tmtextit{(ii)} it is a contraction under the action of positive, and in particular completely positive, trace preserving maps. Consequently, it has been possible to consider the evolution in time of the trace distance between two evolved distinct initial states $\rho_{S}^1 ( 0 )$ and $\rho_{S}^2 ( 0 )$ as an indicator of non-Markovianity, associated to revivals in time of the corresponding trace distance
\begin{displaymath}
  D ( \rho^{1}_{S} ( t ) , \rho^{2}_{S} ( t ) ) -D ( \rho^{1}_{S} ( s ) ,
  \rho^{2}_{S} ( s ) )  > 0,
\end{displaymath}
for some $t\!>\!s$ and a pair $\rho_{S}^1 ( 0 )$ and $\rho_{S}^2 ( 0 )$, as detailed for example in Ref.~{\cite{Breuer2016a}}. Revivals in the trace distance correspond to revivals in distinguishability, i.e. the capability to ascertain the actual initial state by performing measurements on the system only. The information backflow associated to these revivals has been traced back to the establishment of correlations between system and environment as well as to changes in the state of the environment. 

This notion of information backflow can be formalized as follows. Let us identify the total amount of information at time $t$ as the distinguishability of the states of both system and environment
\begin{equation*}
  \mathcal{I}_{\text{tot}} ( t ) =  D ( \rho^{1}_{SE} ( t ) , \rho^{2}_{SE}( t ) ).
\end{equation*}
This quantity is a constant and can be naturally written as the sum of two contributions referring to the information that can be obtained by performing local measurements only, namely $\mathcal{I}_{\text{int}} ( t )\!=\!D (\rho^{1}_{S} ( t ) , \rho^{2}_{S} ( t ) )$, and to the residual information which can only be accessed measuring also the environment $\mathcal{I}_{\text{ext}} ( t ) \!=\!D ( \rho^{1}_{SE} ( t ) , \rho^{2}_{SE} ( t )) -D ( \rho^{1}_{S} ( t ) , \rho^{2}_{S} ( t ) )$. While their sum is a constant, i.e.
\begin{equation*}
\frac{\text{d}}{\text{dt}} \mathcal{I}_{\text{tot}} ( t ) = \frac{\text{d}}{\text{dt}} ( \mathcal{I}_{\text{int}} ( t ) +  \mathcal{I}_{\text{ext}} ( t ) ) =0,
\end{equation*}
nevertheless revivals in the internal information can take place, such that $\mathcal{I}_{\text{int}} ( t ) \!>\! \mathcal{I}_{\text{int}} ( s )$ for $t\!>\!s$, and they are interpreted as information backflow. This interpretation is substantiated by the following inequality
\begin{multline}
   \!\!\!\!\!  D ( \rho^{1}_{S} ( t ) , \rho^{2}_{S} ( t ) ) -D ( \rho^{1}_{S} ( s ) ,
   \rho^{2}_{S} ( s ) ) \leqslant
  D ( \rho^{1}_{E} ( s ) , \rho^{2}_{E} ( s
  ) )   \\
 +D ( \rho^{1}_{SE} ( s ) , \rho^{1}_{S} ( s ) \otimes \rho^{1}_{E} ( s )
  ) +D ( \rho^{2}_{SE} ( s ) , \rho^{2}_{S} ( s ) \otimes \rho^{2}_{E} ( s ) ),
  \label{eq:ineqD}
 \end{multline}
valid for arbitrary $t\! \geqslant \! s$, thus extending previous seminal work on the study of initial correlations~\cite{Laine2010b, Amato2018a}. It appears that to have a local revival of the trace distance at a time $t$ greater than $s$, that is a positive contribution on the l.h.s., at least one of the contributions on the r.h.s., referring to time $s$, has to be positive. In this respect such quantities act as precursors of non-Markovianity, in that their positivity at time $s$ is a necessary, but not sufficient, condition in order to have a larger trace distance at a later time $t$.  Let us further remark that in the case in which the environment is not affected by the system and no correlations are established, corresponding to a perfectly Markovian dynamics, the r.h.s. of the bound is strictly zero, thus forbidding any revival of distinguishability.  Evaluation of Eq.~(\ref{eq:ineqD}) is however quite demanding since it calls for the calculation of the trace distance between both system and environment states, where the latter is in general high-dimensional. Indeed, even the evaluation of the trace distance for the system is a difficult task if one considers higher, possibly infinite, dimensional systems. To overcome these difficulties we will consider a different quantifier of the distinguishability among states and build on the structure of the environment in order to obtain alternative bounds, which still provide necessary conditions for the revivals in distinguishability without calling for measurements on both the system and the whole environment.

In order to consider arbitrary dimensional systems it is natural to introduce as a distance on the state space the so-called Bures distance~{\cite{Fuchs1999a,Vasile2011a}}
\begin{eqnarray}
B ( \rho , \sigma ) & = & \sqrt{2 ( 1-F ( \rho , \sigma ) )} , \nonumber
\end{eqnarray}
where $F ( \rho , \sigma ) = \tmop{Tr} \sqrt{\sqrt{\rho} \sigma \sqrt{\rho}}$ denotes fidelity which is readily computable for both finite dimensional and Gaussian continuous variable systems~\cite{OlivaresEPJST2012, BanchiPRL}. The Bures distance is indeed a metric on the state space, thus satisfying in particular the triangular inequality, and since it is defined in terms of the fidelity, it is a contraction under the action of completely positive trace preserving maps. These properties together with subadditivity with respect to the tensor product allow to reproduce the inequality Eq.~(\ref{eq:ineqD}) in the form
\begin{multline}
  \!\!\!\!\!  B ( \rho^{1}_{S} ( t ) , \rho^{2}_{S} ( t ) ) -B ( \rho^{1}_{S} ( s ) ,
  \rho^{2}_{S} ( s ) ) \leqslant  B ( \rho^{1}_{E} ( s ) , \rho^{2}_{E} ( s
  ) )
  \\
  +B ( \rho^{1}_{SE} ( s ) , \rho^{1}_{S} ( s ) \otimes \rho^{1}_{E} ( s )
  ) +B ( \rho^{2}_{SE} ( s ) , \rho^{2}_{S} ( s ) \otimes \rho^{2}_{E} ( s ) )
  ,  \label{eq:ineqB}
\end{multline}
while retaining its physical interpretation. In this formulation we are, however, still bound to the evaluation of the Bures distance involving all system and environment degrees of freedom, which remains a very difficult task.

We now observe that for the case in which the environment exhibits a natural structuring in terms of constituent subunits (ancillae), as is the case in collision models~\cite{ScaraniPRL2002, BuzekPRA2005}, we can construct a hierarchy of environmental marginals by taking the partial trace with respect to a growing number of environmental ancillae. Exploiting the fact that the partial trace is indeed a completely positive trace preserving transformation, supposing the environment to be composed of $N$ ancillae we have the following chain of bounds for the quantifier of correlations
\begin{multline}
  B ( \tmop{Tr}_{E_{1} , \ldots ,E_{k+1}} \rho^{1}_{SE} ( s ) ,
  \tmop{Tr}_{E_{1} , \ldots ,E_{k+1}}  ( \rho^{1}_{S} ( s ) \otimes
  \rho^{1}_{E} ( s ) ) ) \leqslant
  \\
  B ( \tmop{Tr}_{E_{1} , \ldots ,E_{k}}
  \rho^{1}_{SE} ( s ) , \tmop{Tr}_{E_{1} , \ldots ,E_{k}}  ( \rho^{1}_{S} ( s
  ) \otimes \rho^{1}_{E} ( s ) ) ).  \label{eq:chain1}
\end{multline}
For $k\!=\!0$ we recover the original upper bound $B ( \rho_{SE}^1 (s) , \rho_{S}^1 ( s ) \otimes \rho_{E}^1 ( s ) )$, while for $k\!=\!N$ we have the trivial bound $B ( \rho_{S}^1 ( s ) , \rho_{S}^1 ( s ) )\!=\!0$. Similarly, for the difference in environmental states we have
\begin{multline}
  B ( \tmop{Tr}_{E_{1} , \ldots ,E_{k+1}} \rho^{1}_{E} ( s ) ,
  \tmop{Tr}_{E_{1} , \ldots ,E_{k+1}} \rho^{2}_{E} ( s ) ) \leqslant \\ B (
  \tmop{Tr}_{E_{1} , \ldots ,E_{k}} \rho^{1}_{E} ( s ) , \tmop{Tr}_{E_{1} ,
  \ldots ,E_{k}} \rho^{2}_{E} ( s ) ) .  \label{eq:chain2}
\end{multline}
Thus, it immediately appears that the different terms at the r.h.s. of Eq.~(\ref{eq:ineqB}) can be lower bounded by a whole hierarchy of positive quantities obtained by replacing the environmental states with suitable marginals obtained by tracing out different ancillae. In such a way one can obtain precursors of non-Markovianity which can be more easily evaluated. It is important to stress however that all of these quantities provide lower bounds for the r.h.s., corresponding to quantities which are increasingly easier to evaluate. In this respect, while strict positivity of either of these quantities provides a necessary condition for revivals of Bures distance between distinct pair of initial system states, lower bounding the r.h.s. might lead to a violation of the bound Eq.~\eqref{eq:ineqB}, in the sense that the lower bound for the r.h.s. of Eq.~\eqref{eq:ineqB} is no longer an upper bound to the l.h.s. On the other hand, as we will show in the collision framework in Sec.~\ref{collisionmodel}, it turns out that even the simplest approximation can already lead to a very good estimate of Bures distance revivals. Indeed as shown in Sec.~\ref{validity}, in the collision model framework it appears that already the ``minimal approximation" for the environment, treated as being a single ancilla, actually provides quite reasonable bounds, indicating that the contributions arising from the system interacting with this particular ancilla are the most relevant ones for characterizing the dynamics. This is complementary to Ref.~\cite{CampbellPRA2018} from which we know that the incoming single ancilla is the only really relevant player in dictating the non-Markovianity of the evolution in the corresponding collision model.

We remark that our choice of the Bures distance over other metrics goes beyond its comparative computational simplicity and applicability to infinite dimensional settings already mentioned. The crucial property that any candidate distinguishability measure must have is to be a contraction under the action of completely positive trace preserving maps. While this condition is met for the Bures and trace distances, other norms, such as the Hilbert-Schmidt norm, fail to satisfy it~\cite{WangPRA2009} and thus make them unsuitable choices. Moreover, in order to explicitly relate the bounds to physically meaningful notions, i.e. the established system-environment correlations or environmental changes, one requires a norm so as to exploit the triangular inequality. This requirement therefore rules out quantifiers such as the relative entropy. For these reasons the Bures distance emerges as the most natural metric for the task at hand.

\section{Application to a Collision Model}
\label{collisionmodel}
Let us begin by outlining the basic collision model framework~\cite{ScaraniPRL2002, BuzekPRA2005, McCloskeyPRA2014, StrunzPRA2016, CakmakPRA2017b, CampbellPRA2018, JinNJP2018, EspositoPRX, FrancescoQMQM} which provides a versatile tool for exploring the emergence of non-Markovianity and, due to their construction, serves as the ideal testbed for studying the precursors of non-Markovianity captured in Eq.~\eqref{eq:ineqB} by exploiting Eqs.~\eqref{eq:chain1} and \eqref{eq:chain2}. Following Refs.~\cite{McCloskeyPRA2014, CampbellPRA2018}, the environment is composed of an array of individual ancillae, $E_i$, initially factorized and all with the same initial state. The time evolution is discretized such that at ``time-step" $n$ the system, $S$, collides with ancilla, $E_n$, after which we retain all correlations established by this system-ancilla (SA) interaction while $E_n$ subsequently collides with $E_{n+1}$. At this point we can trace out the degrees of freedom associated with $E_n$, after which we begin time-step $n+1$ where the system collides with $E_{n+1}$ and so on. Due to the intra-environment ancilla-ancilla (AA) collisions, i.e. $E_n$-$E_{n+1}$, $S$ may already share some correlations with $E_{n+1}$ {\it before} they interact~\cite{McCloskeyPRA2014, CampbellPRA2018}. Furthermore, due to the AA collision the incoming state of the ancilla, $E_{n+1}$, that the system interacts with is not typically its initialized state. Thus, the collision model provides a natural setting to explore non-Markovianity arising due to both the establishment of correlations between system and environment together with changes in the state of the environment, as discussed in Sec.~\ref{precursors} and captured succinctly by Eq.~\eqref{eq:ineqB}.

As shown in Ref.~\cite{CampbellPRA2018}, when we restrict to only nearest neighbor AA interactions this corresponds to a memory depth of one and is termed ``first-order Markovian" since we need only concern ourselves with keeping track of the system and one additional ancilla to faithfully simulate of the dynamics. Thus, the evolution follows as
\begin{eqnarray}
\label{model}
\rho_{SE_1E_2}(0) &=& \rho_S(0) \otimes \rho_{E_1}(0) \otimes \rho_{E_2}(0) \nonumber \\
\rho_{SE_1E_2}(1) &=& \mathcal{U}_{E_1 E_{2}} \mathcal{U}_{SE_1} \Big(\rho_{SE_1E_2}(0)\Big) \mathcal{U}_{SE_1}^\dagger \mathcal{U}^\dagger_{E_1 E_{2}} \\
\rho_{SE_2E_3}(2) &=& \mathcal{U}_{E_2 E_{3}} \mathcal{U}_{SE_2} \big(\text{Tr}_{E_1}\left[\rho_{SE_1E_2}(1)\right]\otimes\rho_{E_3}(0)\Big)  \mathcal{U}_{SE_2}^\dagger \mathcal{U}^\dagger_{E_2 E_{3}} \nonumber \\
\vdots \nonumber
\end{eqnarray}
where $\rho_{E_i}(0)$ denotes the initial ancilla state. We exactly recover the standard evolution according to the Markovian master equation when there are no AA collisions, under the condition that the unitary $\mathcal{U}$ is an energy preserving exchange interaction and all constituents have the same free Hamiltonian terms~\cite{McCloskeyPRA2014, PezzuttoNJP}. Despite the framework is independent of the dimensionality of system and ancilla, most studies restrict to the discrete variable qubit states~\cite{McCloskeyPRA2014, StrunzPRA2016, CampbellPRA2018, CakmakPRA2017b}, with only a few exceptions~\cite{JinNJP2018}. Therefore in this work, we are interested in examining any effect that dimensionality may have on the properties of the non-Markovian dynamics, and on the precursors of non-Markovianity, by comparing and contrasting when system and ancilla are discrete variable (DV) qubits with when they are continuous variable (CV) Gaussian states. 

When considering the DV case the interaction will be given by the energy preserving partial swap operation
\begin{equation}
\label{swap}
\mathcal{U}^\text{DV}\equiv\mathcal{S}(\theta) = \cos \theta \openone + i \sin \theta \text{SWAP},
\end{equation} 
with
\begin{equation}
\label{swapmat}
\text{SWAP}=\begin{pmatrix}
 1 & 0 & 0  & 0 \\
 0 & 0  & 1 & 0  \\
 0 & 1 & 0  & 0 \\
 0 & 0 & 0 & 1
\end{pmatrix}.
\end{equation}
We will allow for a generic pure, real initial state of the system, written in the ordered basis $\{ \ket{0} , \ket{1}  \}$,
\begin{equation}
\rho_S(0) =\begin{pmatrix}
\alpha^2 &\alpha\sqrt{1-\alpha^2} \\
\alpha\sqrt{1-\alpha^2} & 1-\alpha^2 
\end{pmatrix}.
\end{equation}
For the CV setting we will assume that SA collisions are described by a beamsplitter
\begin{equation}
\label{beam}
\mathcal{U}^\text{CV-SA} (\theta) = \begin{pmatrix}
 \cos\theta & 0 & \sin\theta  & 0 \\
 0 & \cos\theta  & 0 & \sin\theta  \\
 -\sin\theta & 0 & \cos\theta  & 0 \\
 0 & -\sin\theta & 0 & \cos\theta
\end{pmatrix},
\end{equation}
whereas AA collisions are defined by
 \begin{align}
\label{beamAA}
\mathcal{U}^\text{CV-AA}(\theta)&= \begin{pmatrix}
 \sin\theta & 0 & \cos\theta  & 0 \\
 0 & \sin\theta  & 0 & \cos\theta  \\
 -\cos\theta & 0 & \sin\theta  & 0 \\
 0 & -\cos\theta & 0 & \sin\theta
\end{pmatrix}. \nonumber
\end{align}
We observe that $\mathcal{U}^\text{CV-AA}(\theta) = \mathcal{U}^\text{CV-SA}(\theta-\pi/2)$: the beam splitter implementing the AA interaction has reflectivity and transmittivity inverted w.r.t. the beam splitter realizing the SA interaction. This  corresponds to the shift of a phase factor from the second environmental mode to the first one and allows to avoid a ``jagged''  behaviour of the covariance matrix phases (see~\cite{JinNJP2018}).
We will consider thermal squeezed initial states for the system with covariance matrices given by
\begin{equation}
\label{sigmaini}
\sigma_S(0) = \frac{1+2\nbar}{2} \begin{pmatrix}
\cosh 2r + \sinh 2r &\sinh 2r   \\
\sinh 2r  &\cosh 2r - \sinh 2r
\end{pmatrix}.
\end{equation} 

In what follows we will initialize all environmental ancillae in their ground, respectively vacuum, state for both the DV and CV settings and we will fix the SA (AA) interaction strengths, Eqs.~\eqref{swap} and \eqref{beam}, for both the DV and CV cases to be $\theta=0.05\tfrac{\pi}{2}$ ($\theta=0.9\tfrac{\pi}{2}$). The energy preserving nature of the interaction ensures that the steady-state of system is driven to is exactly the initial state of one of the environmental ancillae, i.e. $\rho_{E_i}(0)$. 

\subsection{Discrete Variable case}
\label{DV}
\begin{figure}[t]
{\bf (a)} \\
\includegraphics[width=0.8\columnwidth]{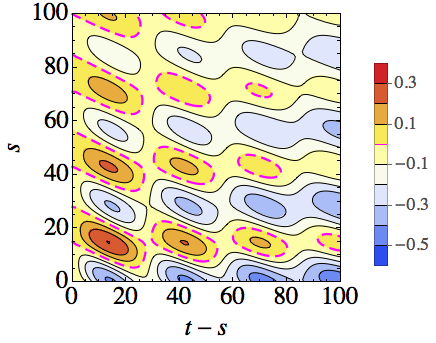}\\
 {\bf (b)} \\
\includegraphics[width=0.7\columnwidth]{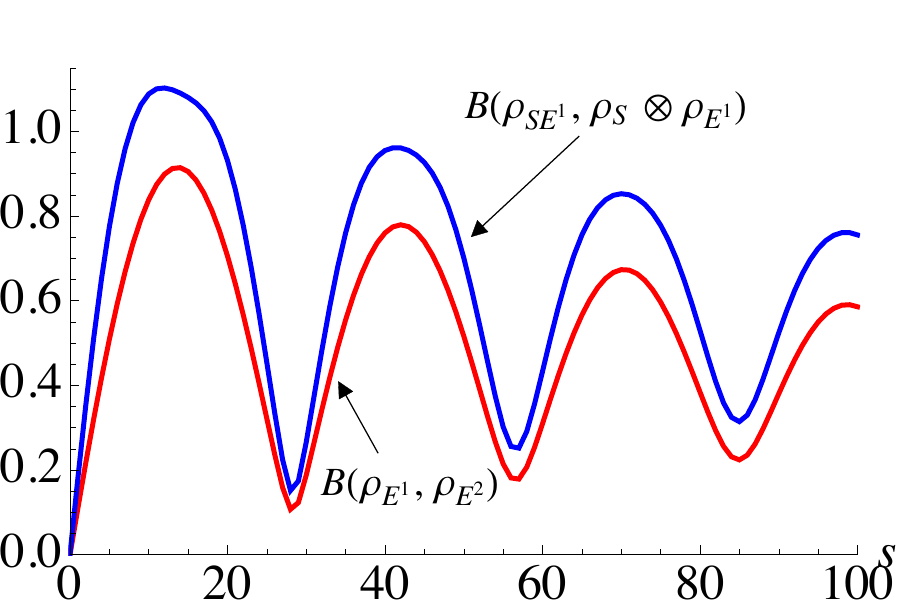}\\
 {\bf (c)}\\
\includegraphics[width=0.85\columnwidth]{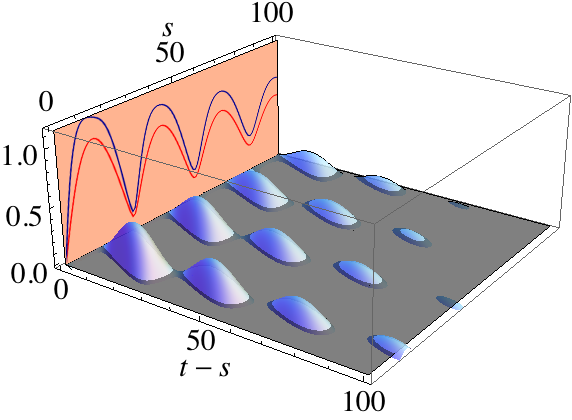}
\caption{Discrete variable (DV) qubit results. We choose as initial system states $\rho_{S}^1\!=\!\ket{1}\!\bra{1}$ and $\rho_{S}^2\!=\!\ket{0}\!\bra{0}$ and all ancillae are initialized in their ground state. {\bf (a)} l.h.s of Eq.~\eqref{eq:ineqB}. The thick, dashed, magenta contour delineates when this quantity is zero. Regions contained within this contour (hotter colours) correspond to revivals and therefore regions of non-Markovianity. {\bf (b)} rhs of Eq.~\eqref{eq:ineqB}. The bottom, red curve corresponds to the first term related to the changes in the environmental state and the top, blue curve is the correlation like term. {\bf (c)} Combined visualization of the previous panels. The gray plane is at zero.}
\label{DVplots}
\end{figure}
For the DV setting we fix the initial state of all ancillae to be $\rho_{E_i}(0)\!=\!\ket{0}\!\bra{0}$ and, in order to evaluate Eq.~\eqref{eq:ineqB}, we consider the two initial system states to be $\rho_{S}^1(0)\!=\!\ket{1}\!\bra{1}$ and $\rho_{S}^2(0)\!=\!\ket{0}\!\bra{0}$. Since $\rho_{S}^2(0)$ is already the steady state of the dynamics, no evolution takes place. This simplifies our evaluation of the various components entering into Eq.~\eqref{eq:ineqB} since the third term on the r.h.s. will be identically zero. Nevertheless we can analyze the non-Markovian dynamics arising in this picture as shown in Fig.~\ref{DVplots}. Panel {\bf (a)} shows the l.h.s of Eq.~\eqref{eq:ineqB}, which accounts only for changes in the system states. As noted previously, we can associate positive values of this quantity with a backflow of information to the environment and therefore periods of non-Markovianity. These periods of non-Markovianity are captured within the dashed, magenta contours. Consider first fixing $s\!=\!0$, we see from the ensuing evolution in $t$ that the l.h.s of Eq.~\eqref{eq:ineqB} remains negative for all $t\!>\!s$, corresponding to the absence of initial correlations. In contrast, if we consider larger values of $s$ we observe that for $t\!>\!s$ revivals can appear. We can clearly see that there is a natural periodicity in these revivals when we witness regions of non-Markovianity in terms of the ``reference-time" $s$. Furthermore, we see the magnitude of the revivals diminishes as $s$ increases, which is to be expected since the state of the system is always driven towards the steady state $\rho_{E_i}(0)$.

We can understand these features by examining our precursors of non-Markovianity. Turning our attention to Fig.~\ref{DVplots} {\bf (b)} we consider the r.h.s of Eq.~\eqref{eq:ineqB} where, by exploiting the hierarchy of bounds in Eqs.~\eqref{eq:chain1} and \eqref{eq:chain2}, we approximate the environment to be a single ancilla, which for a given step is the state of $E_{n+1}$, i.e. the state of the {\it incoming} ancilla after both the SA and AA collisions have taken place. We see that the contributions entering Eq.~\eqref{eq:ineqB} behave qualitatively the same. In particular, focusing on the extremal behaviours we clearly see that when the contributions are largest for a given $s$, this corresponds precisely to when the strongest non-Markovian revivals are present in the l.h.s of Eq.~\eqref{eq:ineqB} for some $t\!>\!s$. We further put this behavior into evidence in panel {\bf (c)} where we combine the data from the preceding panels. 

From these results we can conclude that in the DV case the contributing factors leading to a non-Markovian evolution appear equally important. We can clearly see from panel {\bf (c)} that large revivals in the l.h.s of Eq.~\eqref{eq:ineqB} are associated with both significant amounts of established correlations between system and incoming ancilla and with significant changes to the incoming ancilla state from its initial configuration.

\subsection{Gaussian continuous variable case}
\label{CV}
While considerably less well studied, the case of CV collision models provides an interesting platform for the realization and study of non-Markovianity, particularly in light of the remarkable advances in the manipulation of CV systems for simulating open quantum system dynamics~\cite{SciRepExp, DarwinExp, MataloniExp}. Inline with the DV analysis, we initialize all ancillary modes in the vacuum state. In order to evaluate Eq.~\eqref{eq:ineqB}, we again require two initial system states. Similarly to the previous analysis we fix $\rho_{S}^2$ to be the vacuum state such that no dynamics will occur in this case, thus simplifying the evaluation of Eq.~\eqref{eq:ineqB} since the third term will again be identically zero. At variance with the DV case, we cannot readily initialize our system in a state orthogonal to $\rho_{S}^2$. However, we can consider a strongly squeezed vacuum state, setting $\nbar\!=\!0$ and $r\!=\!0.5$ in Eq.~\eqref{sigmaini}, which is sufficient for our purposes.
\begin{figure}[t]
{\bf (a)} \\
\includegraphics[width=0.8\columnwidth]{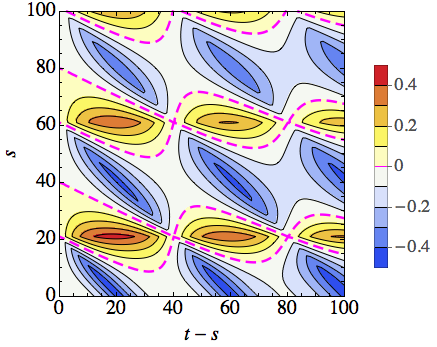}\\
 {\bf (b)}\\
\includegraphics[width=0.7\columnwidth]{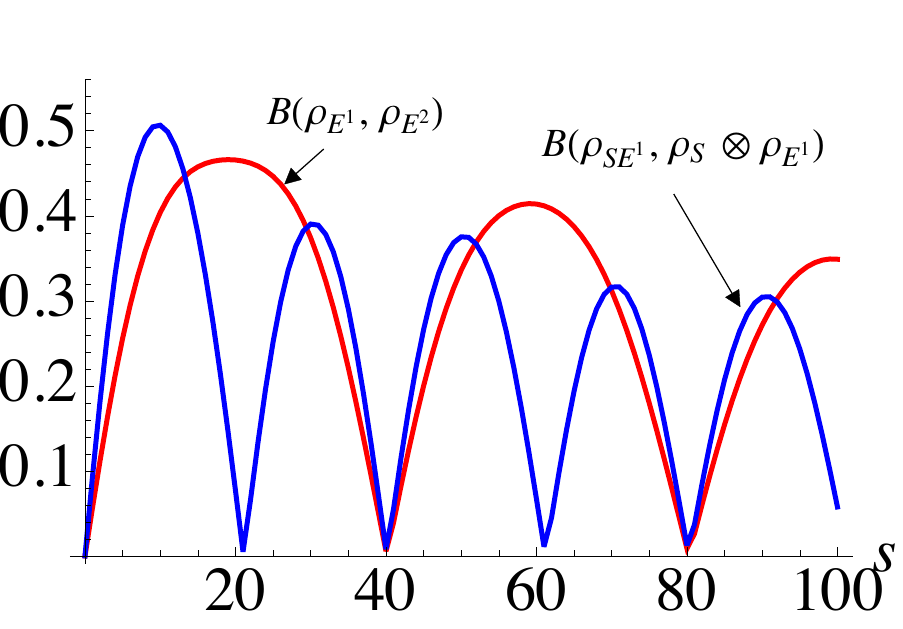}\\
 {\bf (c)}\\
\includegraphics[width=0.8\columnwidth]{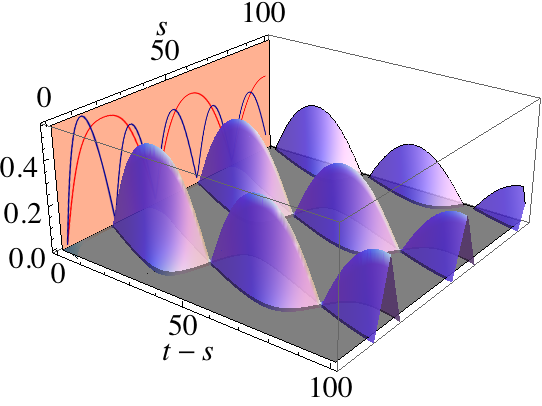}
\caption{Continuous variable (CV) Gaussian results. For initial system states we fix $\nbar\!=\!0$ and $r\!=\!0.5$ in Eq.~\eqref{sigmaini} for $S^1$ while we fix $S^2$ to be the vacuum state. All ancillae are initialized in the vacuum state. {\bf (a)} l.h.s of Eq.~\eqref{eq:ineqB}. The thick, dashed, magenta contour delineates when this quantity is zero. Regions contained within this contour (hot colours) correspond to revivals and therefore regions of non-Markovianity. {\bf (b)} r.h.s of Eq.~\eqref{eq:ineqB}. The red curve corresponds to the first term related to the changes in the environmental state and the blue curve is the correlation like term. {\bf (c)} Combined visualization of the previous panels. The gray plane is at zero.}
\label{CVplots}
\end{figure}

In Fig.~\ref{CVplots} {\bf (a)} we evaluate the l.h.s of Eq.~\eqref{eq:ineqB}. Immediately we can note that there are a number of qualitative similarities with the DV setting. In particular, the clear periodicity in the emergence of regions of non-Markovianity and the decreasing amplitude of the revivals for increasing $s$. However, in contrast with the DV case, we now see that there are special values of $s$ which correspond to continual periods of non-Markovianity. For the considered parameters, fixing $s$ to 20, 60, or 100 collisions we see that the ensuing dynamics is always non-Markovian, while for $s$ equal to 0, 40, or 80 the dynamics is always Markovian. Such a behavior is notably different to the DV case where periods of non-Markovianity are always followed by at least a short period of Markovianity, regardless of the value of $s$. 

This behavior is captured by Fig.~\ref{CVplots} {\bf (b)} where we show the contributions from the r.h.s of Eq.~\eqref{eq:ineqB}. In contrast with the DV case, we now see that the term capturing correlations and the term encompassing environmental changes contribute in a strikingly different manner. In particular, the correlations established between the system and the environment have double the period compared with the changes in the incoming environmental state. This has consequences when we examine the apparent causes of the most non-Markovian regions of Fig.~\ref{CVplots} {\bf (a)}. For the considered parameters, when $s\!=\!20$ we find that for $t\!>\!s$, the l.h.s of Eq.~\eqref{eq:ineqB} achieves its largest revivals. This corresponds to when the first term on the r.h.s. of Eq.~\eqref{eq:ineqB}, related to the change in the incoming environmental state, is at its maximum, while the second term related to the correlations shared between the system and environment is zero.  

A qualitatively similar behavior to the DV case can be seen considering a ``reference-time" $s$ for which the contributions entering into the r.h.s of Eq.~\eqref{eq:ineqB} are large and comparable in magnitude. We again further put into evidence the relation between the precursors of non-Markovianity and the ensuing non-Markovian dynamics in in panel {\bf (c)}.

\subsection{Validity of the Environmental-Size Approximation}
\label{validity}
\begin{figure}[t]
{\bf (a)} \\
\includegraphics[width=0.8\columnwidth]{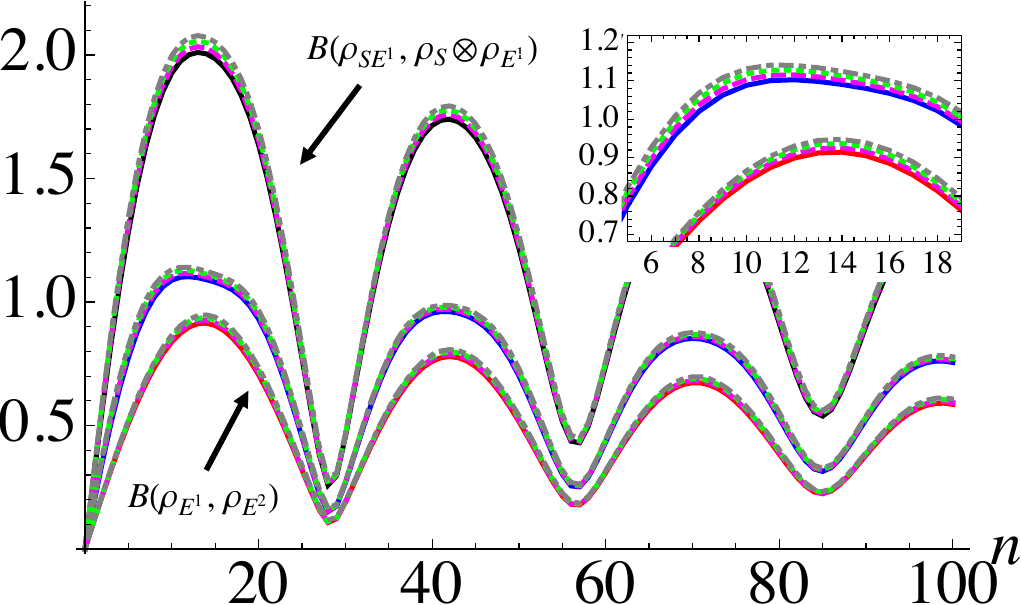}\\
 {\bf (b)}\\
\includegraphics[width=0.8\columnwidth]{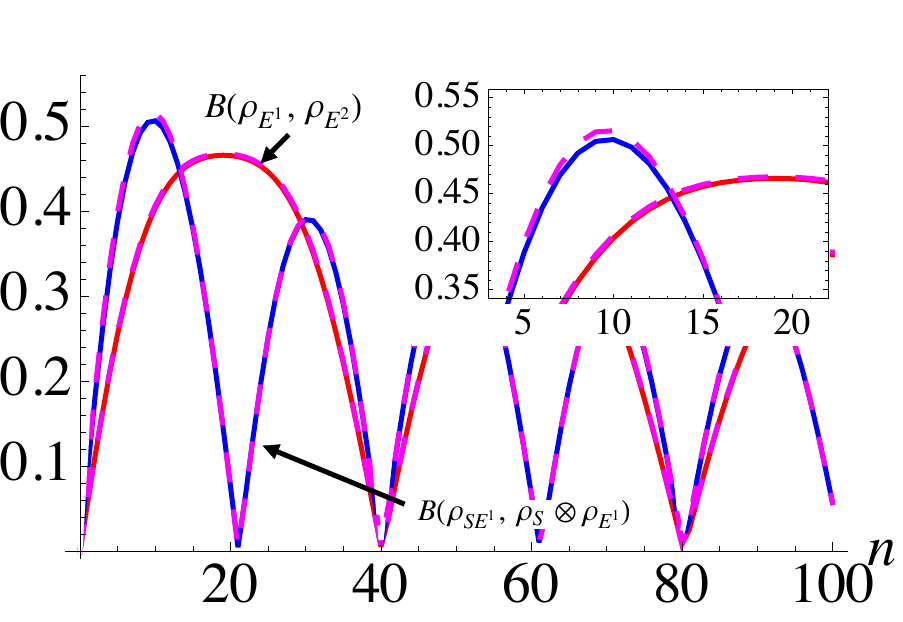}
\caption{We show the r.h.s of Eq.~\eqref{eq:ineqB} for progressively more accurate approximations of the environment. {\bf (a)} DV case where we consider up to 4 ancillae in the evaluation of the bound [from bottom to top]. {\bf (b)} CV case where we approximate the environment as one [lower, solid curves] or two [upper, dashed curves] ancillae in the calculation of the bound. In both panels the insets show a representative zoomed in region, showing the small contribution that storing these additional environmental degrees of freedom provide.}
\label{finitesizeplot}
\end{figure}
In the preceding analyses, to evaluate Eq.~\eqref{eq:ineqB} we made the rather strong assumption that only a single environment ancilla was needed. In particular, we assumed that the incoming ancilla encapsulated all the relevant information to determine the bounds. Here we consider a more careful analysis of the validity of this approximation. To this end we simply examine how the various quantities on the r.h.s of Eq.~\eqref{eq:ineqB} are affected when more environmental degrees of freedom are kept during the simulation. It should be noted that since all quantities are positive, the minimal values observed for only a single ancilla approximation for the environment have already proven to be remarkably good bounds. Indeed, examining Figs.~\ref{DVplots} and \ref{CVplots} we see that even with such an extreme approximation we achieve meaningful and insightful bounds. Furthermore as shown in Ref.~\cite{CampbellPRA2018}, since we know that the dynamics are unaffected by storing more ancillae (and their associated correlations) beyond the memory depth, one might expect that the approximated bounds studied previously are robust. In Fig.~\ref{finitesizeplot} we show that this is indeed the case. The inclusion of more environmental ancillae has only a small effect on the values of the various contributions entering into the r.h.s of Eq.~\eqref{eq:ineqB}. Here we see that the consideration of additional environmental degrees of freedom that are no longer playing an active role in dictating the dynamics of the system provide only a minor contribution to the precursors of non-Markovianity. The seemingly small contribution of these additional correlations can be understood due to the fact that, despite not playing a role in the dynamics, all previously interacted with ancillae share some correlation with the system throughout the entire dynamics~\cite{CampbellPRA2018}. While these correlations appear to be small, they still provide a non-zero contribution to the bounds. These contributions persist until the the system has fully equilibrated with the environment, which corresponds to when the system reaches a factorized state with the environment~\cite{GiovannettiPRA2018}.

\subsection{Non-Markovianity in discrete versus continuous variable models}
\label{DVvsDV}
\begin{figure}[t]
{\bf (a)} \\
\includegraphics[width=0.8\columnwidth]{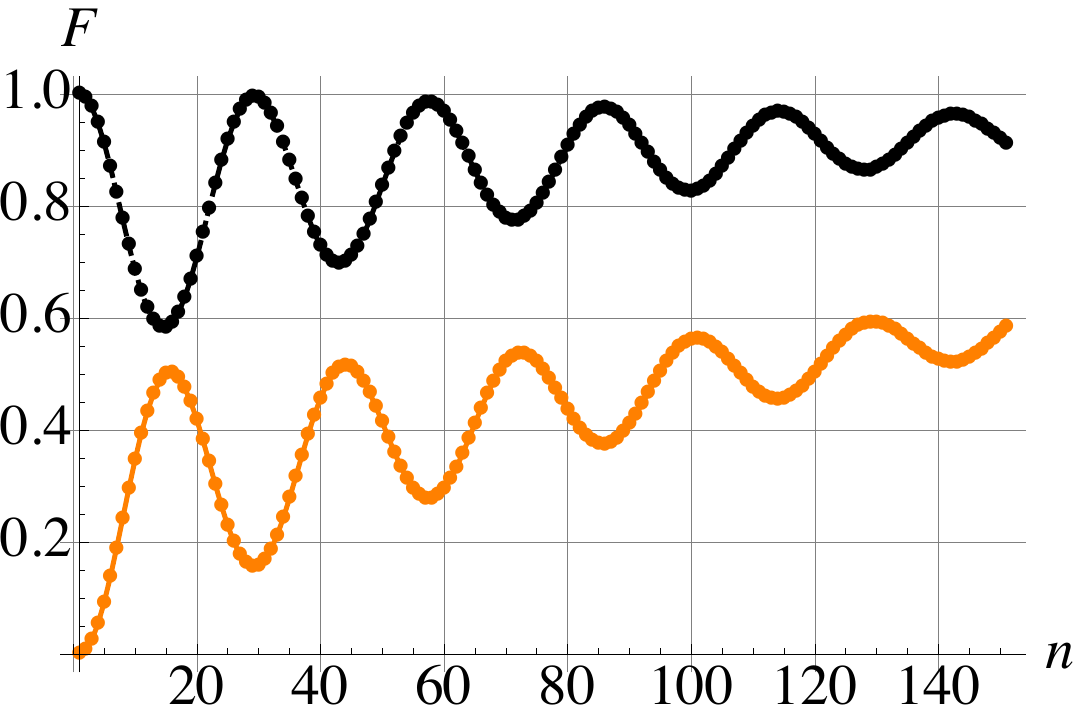}\\
 {\bf (b)}\\
\includegraphics[width=0.8\columnwidth]{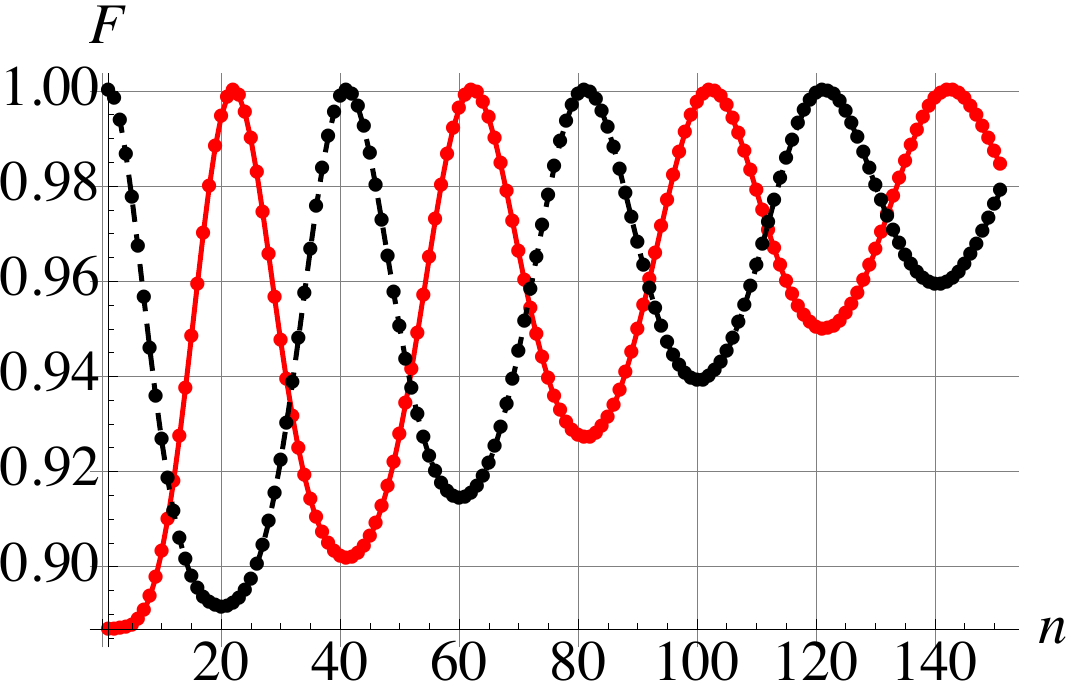}
\caption{Lighter, colored curves show the fidelity between the system state $S^1$ and the steady state. The black curves show the fidelity between the incoming environmental ancilla $E_{n+1}$ and a ``clean" ancilla. {\bf (a)} Discrete variable case. {\bf (b)} Continuous variable case.}
\label{figDVvsCV}
\end{figure}

We have established that while DV and CV settings share several qualitative features, there are notable differences arising. In particular, with regards to the precursors of non-Markovianity we have seen that the various contributions behave quite differently in the two disparate dimensional settings. We can gain a better understanding of the differences between the two settings by considering how close the system gets to the steady state during the dynamics. In Fig.~\ref{figDVvsCV} the lighter, colored curves show the fidelity between the system when it is initialized in $\rho_{S}^1$ with the steady state $\rho_{E_i}(0)$, while the black curves show the fidelity of the incoming ancilla with $\rho_{E_i}(0)$. In panel {\bf (a)} for the DV case we see that the system slowly approaches the steady state, while the incoming ancilla periodically approaches {\it close to} its initial state. In contrast, for the CV case in panel {\bf (b)} we see that the system transients the steady state repeatedly during the dynamics, and similarly the incoming ancilla also periodically returns precisely to its initial state, despite the AA collisions having taken place.

The different behavior of the precursors of non-Markovianity between our considered settings also leads to a final interesting difference. As shown in Ref.~\cite{McCloskeyPRA2014}, a weaker form of non-Markovianity can still be witnessed in the DV collision model if, after the SA collision, the correlations shared between system and ancilla are erased before the AA interaction takes place. In order to see a non-Markovian dynamics however, the initial state of the system must have some coherence, e.g. the system should be initialized in $\ket{+}$ and/or $\ket{-}$. Interestingly, if we consider the same correlation erasure scheme in the CV case, we find that the dynamics is always Markovian, even for strongly squeezed initial system states. 

We remark that, while we presented results referring to zero-temperature environment here, the behaviours described above remain qualitatively the same for different environmental temperatures and different initial states of the system and of the incoming ancillae.

\section{Conclusions}
\label{conclusions}
The description of non-Markovianity of quantum dynamics being due to an information backflow between system and environment leads us to identify the establishment of system-environment correlations and changes in the state of the environment as sources of non-Markovian behavior. In the trace distance approach to non-Markovianity, a bound can be introduced to relate revivals of distinguishability to such changes. We have reformulated this bound in terms of the Bures distance, which is based on the quantum state fidelity. This allows to assess a greater range of physical systems since the fidelity is comparatively easier to compute than the trace distance, in particular when one wishes to consider infinite dimensional systems. Revivals in the l.h.s of the bound, indicating periods of non-Markovianity, can be understood in terms of different contributions due to system-environment correlations and changes in the environmental state at an earlier time, thus establishing the r.h.s of the bound as capturing precursors of non-Markovianity. We have shown that the evaluation of these precursors of non-Markovianity can be simplified when the environment can be decomposed into smaller constituent parts, thus leading to a strategy of general applicability. This is possible by considering a hierarchy of lower bounds to the bound in Bures distance revivals set by the overall amount of established correlations and changes in environmental state. Such lower bounds can be considered whenever the environment exhibits a natural partition within its Hilbert space. We applied this framework to a collision model, where we explored both discrete variable (DV) and continuous variable (CV) settings. Exploiting the considered bound we established that the causes of non-Markovianity in such a collision model showed some qualitative differences between the two disparate dimensional settings, in particular while in the DV case the various contributions to non-Markovianity behaved largely the same, we found that the precursors of non-Markovianity exhibited quite a different behavior in the CV setting. While our results were based on zero-temperature environments, we stress that the same features persist for finite temperature environments and for other choices of initial system states. Our results provide a useful tool for studying the causes of a non-Markovian evolution based on readily computable quantities. Furthermore, we believe our analysis is one of the first to comparatively assess the effect that dimensionality can have on the ensuing non-Markovian character of a given evolution.\bigskip

\acknowledgements 
SC gratefully acknowledges the Science Foundation Ireland Starting Investigator Research Grant ``SpeedDemon" (No. 18/SIRG/5508) for financial support. MP is supported by funding from the European Research Council (ERC) under the European Union's Horizon 2020 research and innovation program (ODYSSEY grant agreement No. 758403). BV gratefully acknowledges support from the Joint Project "Quantum Information Processing in Non- Markovian Quantum Complex Systems" funded by FRIAS/University of Freiburg and IAR/Nagoya University.  BV further acknowledges support from the FFABR project of MIUR and from the UniMi Transition Grant H2020.

\bibliography{CVcollisions}

\begin{thebibliography}{57}%
\makeatletter
\providecommand \@ifxundefined [1]{%
 \@ifx{#1\undefined}
}%
\providecommand \@ifnum [1]{%
 \ifnum #1\expandafter \@firstoftwo
 \else \expandafter \@secondoftwo
 \fi
}%
\providecommand \@ifx [1]{%
 \ifx #1\expandafter \@firstoftwo
 \else \expandafter \@secondoftwo
 \fi
}%
\providecommand \natexlab [1]{#1}%
\providecommand \enquote  [1]{``#1''}%
\providecommand \bibnamefont  [1]{#1}%
\providecommand \bibfnamefont [1]{#1}%
\providecommand \citenamefont [1]{#1}%
\providecommand \href@noop [0]{\@secondoftwo}%
\providecommand \href [0]{\begingroup \@sanitize@url \@href}%
\providecommand \@href[1]{\@@startlink{#1}\@@href}%
\providecommand \@@href[1]{\endgroup#1\@@endlink}%
\providecommand \@sanitize@url [0]{\catcode `\\12\catcode `\$12\catcode
  `\&12\catcode `\#12\catcode `\^12\catcode `\_12\catcode `\%12\relax}%
\providecommand \@@startlink[1]{}%
\providecommand \@@endlink[0]{}%
\providecommand \url  [0]{\begingroup\@sanitize@url \@url }%
\providecommand \@url [1]{\endgroup\@href {#1}{\urlprefix }}%
\providecommand \urlprefix  [0]{URL }%
\providecommand \Eprint [0]{\href }%
\providecommand \doibase [0]{http://dx.doi.org/}%
\providecommand \selectlanguage [0]{\@gobble}%
\providecommand \bibinfo  [0]{\@secondoftwo}%
\providecommand \bibfield  [0]{\@secondoftwo}%
\providecommand \translation [1]{[#1]}%
\providecommand \BibitemOpen [0]{}%
\providecommand \bibitemStop [0]{}%
\providecommand \bibitemNoStop [0]{.\EOS\space}%
\providecommand \EOS [0]{\spacefactor3000\relax}%
\providecommand \BibitemShut  [1]{\csname bibitem#1\endcsname}%
\let\auto@bib@innerbib\@empty
%</preamble>
\bibitem [{\citenamefont {Alicki}\ and\ \citenamefont
  {Lendi}(1987)}]{Alicki1987}%
  \BibitemOpen
  \bibfield  {author} {\bibinfo {author} {\bibfnamefont {R.}~\bibnamefont
  {Alicki}}\ and\ \bibinfo {author} {\bibfnamefont {K.}~\bibnamefont {Lendi}},\
  }\href@noop {} {\emph {\bibinfo {title} {Quantum Dynamical Semigroups and
  Applications}}},\ \bibinfo {series} {Lecture Notes in Physics}, Vol.\
  \bibinfo {volume} {286}\ (\bibinfo  {publisher} {Springer},\ \bibinfo
  {address} {Berlin},\ \bibinfo {year} {1987})\BibitemShut {NoStop}%
\bibitem [{\citenamefont {Weiss}(1993)}]{Weiss1993}%
  \BibitemOpen
  \bibfield  {author} {\bibinfo {author} {\bibfnamefont {U.}~\bibnamefont
  {Weiss}},\ }\href@noop {} {\emph {\bibinfo {title} {Quantum Dissipative
  Systems}}}\ (\bibinfo  {publisher} {World Scientific},\ \bibinfo {address}
  {Singapore},\ \bibinfo {year} {1993})\BibitemShut {NoStop}%
\bibitem [{\citenamefont {Breuer}\ and\ \citenamefont
  {Petruccione}(2002)}]{Breuer2002}%
  \BibitemOpen
  \bibfield  {author} {\bibinfo {author} {\bibfnamefont {H.-P.}\ \bibnamefont
  {Breuer}}\ and\ \bibinfo {author} {\bibfnamefont {F.}~\bibnamefont
  {Petruccione}},\ }\href@noop {} {\emph {\bibinfo {title} {The Theory of Open
  Quantum Systems}}}\ (\bibinfo  {publisher} {Oxford University Press},\
  \bibinfo {address} {Oxford},\ \bibinfo {year} {2002})\BibitemShut {NoStop}%
\bibitem [{\citenamefont {Rivas}\ and\ \citenamefont
  {Huelga}(2012)}]{Rivas2012}%
  \BibitemOpen
  \bibfield  {author} {\bibinfo {author} {\bibfnamefont {A.}~\bibnamefont
  {Rivas}}\ and\ \bibinfo {author} {\bibfnamefont {S.~F.}\ \bibnamefont
  {Huelga}},\ }\href@noop {} {\emph {\bibinfo {title} {Open Quantum Systems: An
  Introduction}}}\ (\bibinfo  {publisher} {Springer},\ \bibinfo {year}
  {2012})\BibitemShut {NoStop}%
\bibitem [{\citenamefont {Breuer}(2012)}]{Breuer2012a}%
  \BibitemOpen
  \bibfield  {author} {\bibinfo {author} {\bibfnamefont {H.-P.}\ \bibnamefont
  {Breuer}},\ }\bibfield  {title} {\enquote {\bibinfo {title} {Foundations and
  measures of quantum non-markovianity},}\ }\href@noop {} {\bibfield  {journal}
  {\bibinfo  {journal} {J. Phys. B}\ }\textbf {\bibinfo {volume} {45}},\
  \bibinfo {pages} {154001} (\bibinfo {year} {2012})}\BibitemShut {NoStop}%
\bibitem [{\citenamefont {Breuer}\ \emph {et~al.}(2016)\citenamefont {Breuer},
  \citenamefont {Laine}, \citenamefont {Piilo},\ and\ \citenamefont
  {Vacchini}}]{Breuer2016a}%
  \BibitemOpen
  \bibfield  {author} {\bibinfo {author} {\bibfnamefont {H.-P.}\ \bibnamefont
  {Breuer}}, \bibinfo {author} {\bibfnamefont {E.-M.}\ \bibnamefont {Laine}},
  \bibinfo {author} {\bibfnamefont {J.}~\bibnamefont {Piilo}}, \ and\ \bibinfo
  {author} {\bibfnamefont {B.}~\bibnamefont {Vacchini}},\ }\bibfield  {title}
  {\enquote {\bibinfo {title} {\textit{Colloquium} : Non-markovian dynamics in
  open quantum systems},}\ }\href {\doibase 10.1103/RevModPhys.88.021002}
  {\bibfield  {journal} {\bibinfo  {journal} {Rev. Mod. Phys.}\ }\textbf
  {\bibinfo {volume} {88}},\ \bibinfo {pages} {021002} (\bibinfo {year}
  {2016})}\BibitemShut {NoStop}%
\bibitem [{\citenamefont {Rivas}\ \emph {et~al.}(2014)\citenamefont {Rivas},
  \citenamefont {Huelga},\ and\ \citenamefont {Plenio}}]{Rivas2014a}%
  \BibitemOpen
  \bibfield  {author} {\bibinfo {author} {\bibfnamefont {A.}~\bibnamefont
  {Rivas}}, \bibinfo {author} {\bibfnamefont {S.~F.}\ \bibnamefont {Huelga}}, \
  and\ \bibinfo {author} {\bibfnamefont {M.~B.}\ \bibnamefont {Plenio}},\
  }\bibfield  {title} {\enquote {\bibinfo {title} {Quantum non-markovianity:
  Characterization, quantification and detection},}\ }\href {\doibase
  10.1088/0034-4885/77/9/094001} {\bibfield  {journal} {\bibinfo  {journal}
  {Rep. Prog. Phys.}\ }\textbf {\bibinfo {volume} {77}},\ \bibinfo {pages}
  {094001} (\bibinfo {year} {2014})}\BibitemShut {NoStop}%
\bibitem [{\citenamefont {de~Vega}\ and\ \citenamefont
  {Alonso}(2017)}]{Devega2017a}%
  \BibitemOpen
  \bibfield  {author} {\bibinfo {author} {\bibfnamefont {I.}~\bibnamefont
  {de~Vega}}\ and\ \bibinfo {author} {\bibfnamefont {D.}~\bibnamefont
  {Alonso}},\ }\bibfield  {title} {\enquote {\bibinfo {title} {Dynamics of
  non-markovian open quantum systems},}\ }\href {\doibase
  10.1103/RevModPhys.89.015001} {\bibfield  {journal} {\bibinfo  {journal}
  {Rev. Mod. Phys.}\ }\textbf {\bibinfo {volume} {89}},\ \bibinfo {pages}
  {015001} (\bibinfo {year} {2017})}\BibitemShut {NoStop}%
\bibitem [{\citenamefont {Li}\ \emph {et~al.}(2018)\citenamefont {Li},
  \citenamefont {Hall},\ and\ \citenamefont {Wiseman}}]{Li2018a}%
  \BibitemOpen
  \bibfield  {author} {\bibinfo {author} {\bibfnamefont {L.}~\bibnamefont
  {Li}}, \bibinfo {author} {\bibfnamefont {M.~J.~W.}\ \bibnamefont {Hall}}, \
  and\ \bibinfo {author} {\bibfnamefont {H.~M.}\ \bibnamefont {Wiseman}},\
  }\bibfield  {title} {\enquote {\bibinfo {title} {Concepts of quantum
  non-markovianity: A hierarchy},}\ }\href {\doibase
  https://doi.org/10.1016/j.physrep.2018.07.001} {\bibfield  {journal}
  {\bibinfo  {journal} {Physics Reports}\ }\textbf {\bibinfo {volume} {759}},\
  \bibinfo {pages} {1 -- 51} (\bibinfo {year} {2018})}\BibitemShut {NoStop}%
\bibitem [{\citenamefont {Huelga}\ \emph {et~al.}(2012)\citenamefont {Huelga},
  \citenamefont {Rivas},\ and\ \citenamefont {Plenio}}]{Plenio1}%
  \BibitemOpen
  \bibfield  {author} {\bibinfo {author} {\bibfnamefont {S.~F.}\ \bibnamefont
  {Huelga}}, \bibinfo {author} {\bibfnamefont {\'A.}\ \bibnamefont {Rivas}}, \
  and\ \bibinfo {author} {\bibfnamefont {M.~B.}\ \bibnamefont {Plenio}},\
  }\bibfield  {title} {\enquote {\bibinfo {title} {Non-markovianity-assisted
  steady state entanglement},}\ }\href {\doibase
  10.1103/PhysRevLett.108.160402} {\bibfield  {journal} {\bibinfo  {journal}
  {Phys. Rev. Lett.}\ }\textbf {\bibinfo {volume} {108}},\ \bibinfo {pages}
  {160402} (\bibinfo {year} {2012})}\BibitemShut {NoStop}%
\bibitem [{\citenamefont {Chin}\ \emph {et~al.}(2012)\citenamefont {Chin},
  \citenamefont {Huelga},\ and\ \citenamefont {Plenio}}]{Plenio2}%
  \BibitemOpen
  \bibfield  {author} {\bibinfo {author} {\bibfnamefont {A.~W.}\ \bibnamefont
  {Chin}}, \bibinfo {author} {\bibfnamefont {S.~F.}\ \bibnamefont {Huelga}}, \
  and\ \bibinfo {author} {\bibfnamefont {M.~B.}\ \bibnamefont {Plenio}},\
  }\bibfield  {title} {\enquote {\bibinfo {title} {Quantum metrology in
  non-markovian environments},}\ }\href {\doibase
  10.1103/PhysRevLett.109.233601} {\bibfield  {journal} {\bibinfo  {journal}
  {Phys. Rev. Lett.}\ }\textbf {\bibinfo {volume} {109}},\ \bibinfo {pages}
  {233601} (\bibinfo {year} {2012})}\BibitemShut {NoStop}%
\bibitem [{\citenamefont {Cimmarusti}\ \emph {et~al.}(2015)\citenamefont
  {Cimmarusti}, \citenamefont {Yan}, \citenamefont {Patterson}, \citenamefont
  {Corcos}, \citenamefont {Orozco},\ and\ \citenamefont
  {Deffner}}]{DeffnerPRL2015}%
  \BibitemOpen
  \bibfield  {author} {\bibinfo {author} {\bibfnamefont {A.~D.}\ \bibnamefont
  {Cimmarusti}}, \bibinfo {author} {\bibfnamefont {Z.}~\bibnamefont {Yan}},
  \bibinfo {author} {\bibfnamefont {B.~D.}\ \bibnamefont {Patterson}}, \bibinfo
  {author} {\bibfnamefont {L.~P.}\ \bibnamefont {Corcos}}, \bibinfo {author}
  {\bibfnamefont {L.~A.}\ \bibnamefont {Orozco}}, \ and\ \bibinfo {author}
  {\bibfnamefont {S.}~\bibnamefont {Deffner}},\ }\bibfield  {title} {\enquote
  {\bibinfo {title} {Environment-assisted speed-up of the field evolution in
  cavity quantum electrodynamics},}\ }\href {\doibase
  10.1103/PhysRevLett.114.233602} {\bibfield  {journal} {\bibinfo  {journal}
  {Phys. Rev. Lett.}\ }\textbf {\bibinfo {volume} {114}},\ \bibinfo {pages}
  {233602} (\bibinfo {year} {2015})}\BibitemShut {NoStop}%
\bibitem [{\citenamefont {{Pezzutto}}\ \emph {et~al.}(2019)\citenamefont
  {{Pezzutto}}, \citenamefont {{Paternostro}},\ and\ \citenamefont
  {{Omar}}}]{PezzuttoQST}%
  \BibitemOpen
  \bibfield  {author} {\bibinfo {author} {\bibfnamefont {M.}~\bibnamefont
  {{Pezzutto}}}, \bibinfo {author} {\bibfnamefont {M.}~\bibnamefont
  {{Paternostro}}}, \ and\ \bibinfo {author} {\bibfnamefont {Y.}~\bibnamefont
  {{Omar}}},\ }\bibfield  {title} {\enquote {\bibinfo {title} {{An
  out-of-equilibrium non-Markovian Quantum Heat Engine}},}\ }\href {\doibase
  10.1088/2058-9565/aaf5b4} {\bibfield  {journal} {\bibinfo  {journal} {Quantum
  Sci. Technol.}\ }\textbf {\bibinfo {volume} {4}},\ \bibinfo {pages} {025002}
  (\bibinfo {year} {2019})}\BibitemShut {NoStop}%
\bibitem [{\citenamefont {\ifmmode~\mbox{\c{C}}\else \c{C}\fi{}akmak}\ \emph
  {et~al.}(2019)\citenamefont {\ifmmode~\mbox{\c{C}}\else \c{C}\fi{}akmak},
  \citenamefont {Campbell}, \citenamefont {Vacchini}, \citenamefont
  {M\"ustecapl\ifmmode \imath \else \i \fi{}o\ifmmode~\breve{g}\else
  \u{g}\fi{}lu},\ and\ \citenamefont {Paternostro}}]{BarisArXiv}%
  \BibitemOpen
  \bibfield  {author} {\bibinfo {author} {\bibfnamefont {B.}~\bibnamefont
  {\ifmmode~\mbox{\c{C}}\else \c{C}\fi{}akmak}}, \bibinfo {author}
  {\bibfnamefont {S.}~\bibnamefont {Campbell}}, \bibinfo {author}
  {\bibfnamefont {B.}~\bibnamefont {Vacchini}}, \bibinfo {author}
  {\bibfnamefont {\"O.~E.}\ \bibnamefont {M\"ustecapl\ifmmode \imath \else \i
  \fi{}o\ifmmode~\breve{g}\else \u{g}\fi{}lu}}, \ and\ \bibinfo {author}
  {\bibfnamefont {M.}~\bibnamefont {Paternostro}},\ }\bibfield  {title}
  {\enquote {\bibinfo {title} {Robust multipartite entanglement generation via
  a collision model},}\ }\href {\doibase 10.1103/PhysRevA.99.012319} {\bibfield
   {journal} {\bibinfo  {journal} {Phys. Rev. A}\ }\textbf {\bibinfo {volume}
  {99}},\ \bibinfo {pages} {012319} (\bibinfo {year} {2019})}\BibitemShut
  {NoStop}%
\bibitem [{\citenamefont {Sakuldee}\ \emph {et~al.}(2018)\citenamefont
  {Sakuldee}, \citenamefont {Milz}, \citenamefont {Pollock},\ and\
  \citenamefont {Modi}}]{ModiJPhysA}%
  \BibitemOpen
  \bibfield  {author} {\bibinfo {author} {\bibfnamefont {F.}~\bibnamefont
  {Sakuldee}}, \bibinfo {author} {\bibfnamefont {S.}~\bibnamefont {Milz}},
  \bibinfo {author} {\bibfnamefont {F.~A.}\ \bibnamefont {Pollock}}, \ and\
  \bibinfo {author} {\bibfnamefont {K.}~\bibnamefont {Modi}},\ }\bibfield
  {title} {\enquote {\bibinfo {title} {Non-markovian quantum control as
  coherent stochastic trajectories},}\ }\href {\doibase
  10.1088/1751-8121/aabb1e} {\bibfield  {journal} {\bibinfo  {journal} {J.
  Phys. A: Math. Theor.}\ }\textbf {\bibinfo {volume} {51}},\ \bibinfo {pages}
  {414014} (\bibinfo {year} {2018})}\BibitemShut {NoStop}%
\bibitem [{\citenamefont {Dong}\ \emph {et~al.}(2018)\citenamefont {Dong},
  \citenamefont {Zheng}, \citenamefont {Li}, \citenamefont {Li}, \citenamefont
  {Chen}, \citenamefont {Guo},\ and\ \citenamefont {Sun}}]{NJPQI}%
  \BibitemOpen
  \bibfield  {author} {\bibinfo {author} {\bibfnamefont {Y.}~\bibnamefont
  {Dong}}, \bibinfo {author} {\bibfnamefont {Y.}~\bibnamefont {Zheng}},
  \bibinfo {author} {\bibfnamefont {S.}~\bibnamefont {Li}}, \bibinfo {author}
  {\bibfnamefont {C.-C.}\ \bibnamefont {Li}}, \bibinfo {author} {\bibfnamefont
  {X.-D.}\ \bibnamefont {Chen}}, \bibinfo {author} {\bibfnamefont {G.-C.}\
  \bibnamefont {Guo}}, \ and\ \bibinfo {author} {\bibfnamefont {F.-W.}\
  \bibnamefont {Sun}},\ }\bibfield  {title} {\enquote {\bibinfo {title}
  {Non-markovianity-assisted high-fidelity deutsch-jozsa algorithm in
  diamond},}\ }\href {\doibase 10.1038/s41534-017-0053-z} {\bibfield  {journal}
  {\bibinfo  {journal} {npj Quantum Inf.}\ }\textbf {\bibinfo {volume} {4}},\
  \bibinfo {pages} {3} (\bibinfo {year} {2018})}\BibitemShut {NoStop}%
\bibitem [{\citenamefont {Pollock}\ \emph {et~al.}(2018)\citenamefont
  {Pollock}, \citenamefont {Rodriguez-Rosario}, \citenamefont {Frauenheim},
  \citenamefont {Paternostro},\ and\ \citenamefont {Modi}}]{Pollock2018a}%
  \BibitemOpen
  \bibfield  {author} {\bibinfo {author} {\bibfnamefont {F.~A.}\ \bibnamefont
  {Pollock}}, \bibinfo {author} {\bibfnamefont {C.}~\bibnamefont
  {Rodriguez-Rosario}}, \bibinfo {author} {\bibfnamefont {T.}~\bibnamefont
  {Frauenheim}}, \bibinfo {author} {\bibfnamefont {M.}~\bibnamefont
  {Paternostro}}, \ and\ \bibinfo {author} {\bibfnamefont {K.}~\bibnamefont
  {Modi}},\ }\bibfield  {title} {\enquote {\bibinfo {title} {Operational markov
  condition for quantum processes},}\ }\href {\doibase
  10.1103/PhysRevLett.120.040405} {\bibfield  {journal} {\bibinfo  {journal}
  {Phys. Rev. Lett.}\ }\textbf {\bibinfo {volume} {120}},\ \bibinfo {pages}
  {040405} (\bibinfo {year} {2018})}\BibitemShut {NoStop}%
\bibitem [{\citenamefont {Taranto}\ \emph {et~al.}(2019)\citenamefont
  {Taranto}, \citenamefont {Milz}, \citenamefont {Pollock},\ and\ \citenamefont
  {Modi}}]{Modiarxiv}%
  \BibitemOpen
  \bibfield  {author} {\bibinfo {author} {\bibfnamefont {P.}~\bibnamefont
  {Taranto}}, \bibinfo {author} {\bibfnamefont {S.}~\bibnamefont {Milz}},
  \bibinfo {author} {\bibfnamefont {F.~A.}\ \bibnamefont {Pollock}}, \ and\
  \bibinfo {author} {\bibfnamefont {K.}~\bibnamefont {Modi}},\ }\bibfield
  {title} {\enquote {\bibinfo {title} {Structure of quantum stochastic
  processes with finite markov order},}\ }\href {\doibase
  10.1103/PhysRevA.99.042108} {\bibfield  {journal} {\bibinfo  {journal} {Phys.
  Rev. A}\ }\textbf {\bibinfo {volume} {99}},\ \bibinfo {pages} {042108}
  (\bibinfo {year} {2019})}\BibitemShut {NoStop}%
\bibitem [{\citenamefont {Chruscinski}\ \emph {et~al.}(2011)\citenamefont
  {Chruscinski}, \citenamefont {Kossakowski},\ and\ \citenamefont
  {Rivas}}]{Chruscinski2011a}%
  \BibitemOpen
  \bibfield  {author} {\bibinfo {author} {\bibfnamefont {D.}~\bibnamefont
  {Chruscinski}}, \bibinfo {author} {\bibfnamefont {A.}~\bibnamefont
  {Kossakowski}}, \ and\ \bibinfo {author} {\bibfnamefont {A.}~\bibnamefont
  {Rivas}},\ }\bibfield  {title} {\enquote {\bibinfo {title} {On measures of
  non-markovianity: divisibility vs. backflow of information},}\ }\href
  {\doibase 10.1103/PhysRevA.83.052128} {\bibfield  {journal} {\bibinfo
  {journal} {Phys. Rev. A}\ }\textbf {\bibinfo {volume} {83}},\ \bibinfo
  {pages} {052128} (\bibinfo {year} {2011})}\BibitemShut {NoStop}%
\bibitem [{\citenamefont {Wi\ss{}mann}\ \emph {et~al.}(2015)\citenamefont
  {Wi\ss{}mann}, \citenamefont {Breuer},\ and\ \citenamefont
  {Vacchini}}]{Wissmann2015a}%
  \BibitemOpen
  \bibfield  {author} {\bibinfo {author} {\bibfnamefont {S.}~\bibnamefont
  {Wi\ss{}mann}}, \bibinfo {author} {\bibfnamefont {H.-P.}\ \bibnamefont
  {Breuer}}, \ and\ \bibinfo {author} {\bibfnamefont {B.}~\bibnamefont
  {Vacchini}},\ }\bibfield  {title} {\enquote {\bibinfo {title} {Generalized
  trace-distance measure connecting quantum and classical non-markovianity},}\
  }\href {\doibase 10.1103/PhysRevA.92.042108} {\bibfield  {journal} {\bibinfo
  {journal} {Phys. Rev. A}\ }\textbf {\bibinfo {volume} {92}},\ \bibinfo
  {pages} {042108} (\bibinfo {year} {2015})}\BibitemShut {NoStop}%
\bibitem [{\citenamefont {Breuer}\ \emph {et~al.}(2018)\citenamefont {Breuer},
  \citenamefont {Amato},\ and\ \citenamefont {Vacchini}}]{Breuer2018a}%
  \BibitemOpen
  \bibfield  {author} {\bibinfo {author} {\bibfnamefont {H.-P.}\ \bibnamefont
  {Breuer}}, \bibinfo {author} {\bibfnamefont {G.}~\bibnamefont {Amato}}, \
  and\ \bibinfo {author} {\bibfnamefont {B.}~\bibnamefont {Vacchini}},\
  }\bibfield  {title} {\enquote {\bibinfo {title} {Mixing-induced quantum
  non-markovianity and information flow},}\ }\href
  {http://stacks.iop.org/1367-2630/20/i=4/a=043007} {\bibfield  {journal}
  {\bibinfo  {journal} {New J. Phys.}\ }\textbf {\bibinfo {volume} {20}},\
  \bibinfo {pages} {043007} (\bibinfo {year} {2018})}\BibitemShut {NoStop}%
\bibitem [{\citenamefont {Chruscinski}\ \emph {et~al.}(2018)\citenamefont
  {Chruscinski}, \citenamefont {Rivas},\ and\ \citenamefont
  {St\o{}rmer}}]{Chruscinski2018a}%
  \BibitemOpen
  \bibfield  {author} {\bibinfo {author} {\bibfnamefont {D.}~\bibnamefont
  {Chruscinski}}, \bibinfo {author} {\bibfnamefont {\'A.}\ \bibnamefont
  {Rivas}}, \ and\ \bibinfo {author} {\bibfnamefont {E.}~\bibnamefont
  {St\o{}rmer}},\ }\bibfield  {title} {\enquote {\bibinfo {title} {Divisibility
  and information flow notions of quantum markovianity for noninvertible
  dynamical maps},}\ }\href {\doibase 10.1103/PhysRevLett.121.080407}
  {\bibfield  {journal} {\bibinfo  {journal} {Phys. Rev. Lett.}\ }\textbf
  {\bibinfo {volume} {121}},\ \bibinfo {pages} {080407} (\bibinfo {year}
  {2018})}\BibitemShut {NoStop}%
\bibitem [{\citenamefont {Breuer}\ \emph {et~al.}(2009)\citenamefont {Breuer},
  \citenamefont {Laine},\ and\ \citenamefont {Piilo}}]{BreuerPRL2009}%
  \BibitemOpen
  \bibfield  {author} {\bibinfo {author} {\bibfnamefont {H.-P.}\ \bibnamefont
  {Breuer}}, \bibinfo {author} {\bibfnamefont {E.-M.}\ \bibnamefont {Laine}}, \
  and\ \bibinfo {author} {\bibfnamefont {J.}~\bibnamefont {Piilo}},\ }\bibfield
   {title} {\enquote {\bibinfo {title} {Measure for the degree of non-markovian
  behavior of quantum processes in open systems},}\ }\href {\doibase
  10.1103/PhysRevLett.103.210401} {\bibfield  {journal} {\bibinfo  {journal}
  {Phys. Rev. Lett.}\ }\textbf {\bibinfo {volume} {103}},\ \bibinfo {pages}
  {210401} (\bibinfo {year} {2009})}\BibitemShut {NoStop}%
\bibitem [{\citenamefont {Laine}\ \emph {et~al.}(2010)\citenamefont {Laine},
  \citenamefont {Piilo},\ and\ \citenamefont {Breuer}}]{Laine2010b}%
  \BibitemOpen
  \bibfield  {author} {\bibinfo {author} {\bibfnamefont {E.-M.}\ \bibnamefont
  {Laine}}, \bibinfo {author} {\bibfnamefont {J.}~\bibnamefont {Piilo}}, \ and\
  \bibinfo {author} {\bibfnamefont {H.-P.}\ \bibnamefont {Breuer}},\ }\bibfield
   {title} {\enquote {\bibinfo {title} {Witness for initial system-environment
  correlations in open system dynamics},}\ }\href {\doibase
  10.1209/0295-5075/92/60010} {\bibfield  {journal} {\bibinfo  {journal} {EPL}\
  }\textbf {\bibinfo {volume} {92}},\ \bibinfo {pages} {60010} (\bibinfo {year}
  {2010})}\BibitemShut {NoStop}%
\bibitem [{\citenamefont {Amato}\ \emph {et~al.}(2018)\citenamefont {Amato},
  \citenamefont {Breuer},\ and\ \citenamefont {Vacchini}}]{Amato2018a}%
  \BibitemOpen
  \bibfield  {author} {\bibinfo {author} {\bibfnamefont {G.}~\bibnamefont
  {Amato}}, \bibinfo {author} {\bibfnamefont {H.-P.}\ \bibnamefont {Breuer}}, \
  and\ \bibinfo {author} {\bibfnamefont {B.}~\bibnamefont {Vacchini}},\
  }\bibfield  {title} {\enquote {\bibinfo {title} {Generalized trace distance
  approach to quantum non-markovianity and detection of initial
  correlations},}\ }\href {\doibase 10.1103/PhysRevA.98.012120} {\bibfield
  {journal} {\bibinfo  {journal} {Phys. Rev. A}\ }\textbf {\bibinfo {volume}
  {98}},\ \bibinfo {pages} {012120} (\bibinfo {year} {2018})}\BibitemShut
  {NoStop}%
\bibitem [{\citenamefont {Mazzola}\ \emph {et~al.}(2012)\citenamefont
  {Mazzola}, \citenamefont {Rodr\'iguez-Rosario}, \citenamefont {Modi},\ and\
  \citenamefont {Paternostro}}]{Mazzola2012a}%
  \BibitemOpen
  \bibfield  {author} {\bibinfo {author} {\bibfnamefont {L.}~\bibnamefont
  {Mazzola}}, \bibinfo {author} {\bibfnamefont {C.~A.}\ \bibnamefont
  {Rodr\'iguez-Rosario}}, \bibinfo {author} {\bibfnamefont {K.}~\bibnamefont
  {Modi}}, \ and\ \bibinfo {author} {\bibfnamefont {M.}~\bibnamefont
  {Paternostro}},\ }\bibfield  {title} {\enquote {\bibinfo {title} {Dynamical
  role of system-environment correlations in non-markovian dynamics},}\ }\href
  {\doibase 10.1103/PhysRevA.86.010102} {\bibfield  {journal} {\bibinfo
  {journal} {Phys. Rev. A}\ }\textbf {\bibinfo {volume} {86}},\ \bibinfo
  {pages} {010102} (\bibinfo {year} {2012})}\BibitemShut {NoStop}%
\bibitem [{\citenamefont {Rodr\'iguez-Rosario}\ \emph
  {et~al.}(2012)\citenamefont {Rodr\'iguez-Rosario}, \citenamefont {Modi},
  \citenamefont {Mazzola},\ and\ \citenamefont
  {Aspuru-Guzik}}]{Rodriguez2012a}%
  \BibitemOpen
  \bibfield  {author} {\bibinfo {author} {\bibfnamefont {C.~A}\ \bibnamefont
  {Rodr\'iguez-Rosario}}, \bibinfo {author} {\bibfnamefont {K.}~\bibnamefont
  {Modi}}, \bibinfo {author} {\bibfnamefont {L.}~\bibnamefont {Mazzola}}, \
  and\ \bibinfo {author} {\bibfnamefont {A.}~\bibnamefont {Aspuru-Guzik}},\
  }\bibfield  {title} {\enquote {\bibinfo {title} {Unification of witnessing
  initial system-environment correlations and witnessing non-markovianity},}\
  }\href {\doibase 10.1209/0295-5075/99/20010} {\bibfield  {journal} {\bibinfo
  {journal} {EPL}\ }\textbf {\bibinfo {volume} {99}},\ \bibinfo {pages} {20010}
  (\bibinfo {year} {2012})}\BibitemShut {NoStop}%
\bibitem [{\citenamefont {Smirne}\ \emph {et~al.}(2013)\citenamefont {Smirne},
  \citenamefont {Mazzola}, \citenamefont {Paternostro},\ and\ \citenamefont
  {Vacchini}}]{Smirne2013b}%
  \BibitemOpen
  \bibfield  {author} {\bibinfo {author} {\bibfnamefont {A.}~\bibnamefont
  {Smirne}}, \bibinfo {author} {\bibfnamefont {L.}~\bibnamefont {Mazzola}},
  \bibinfo {author} {\bibfnamefont {M.}~\bibnamefont {Paternostro}}, \ and\
  \bibinfo {author} {\bibfnamefont {B.}~\bibnamefont {Vacchini}},\ }\bibfield
  {title} {\enquote {\bibinfo {title} {Interaction-induced correlations and
  non-markovianity of quantum dynamics},}\ }\href {\doibase
  10.1103/PhysRevA.87.052129} {\bibfield  {journal} {\bibinfo  {journal} {Phys.
  Rev. A}\ }\textbf {\bibinfo {volume} {87}},\ \bibinfo {pages} {052129}
  (\bibinfo {year} {2013})}\BibitemShut {NoStop}%
\bibitem [{\citenamefont {Iles-Smith}\ \emph {et~al.}(2014)\citenamefont
  {Iles-Smith}, \citenamefont {Lambert},\ and\ \citenamefont
  {Nazir}}]{NazirPRA2014}%
  \BibitemOpen
  \bibfield  {author} {\bibinfo {author} {\bibfnamefont {J.}~\bibnamefont
  {Iles-Smith}}, \bibinfo {author} {\bibfnamefont {N.}~\bibnamefont {Lambert}},
  \ and\ \bibinfo {author} {\bibfnamefont {A.}~\bibnamefont {Nazir}},\
  }\bibfield  {title} {\enquote {\bibinfo {title} {Environmental dynamics,
  correlations, and the emergence of noncanonical equilibrium states in open
  quantum systems},}\ }\href {\doibase 10.1103/PhysRevA.90.032114} {\bibfield
  {journal} {\bibinfo  {journal} {Phys. Rev. A}\ }\textbf {\bibinfo {volume}
  {90}},\ \bibinfo {pages} {032114} (\bibinfo {year} {2014})}\BibitemShut
  {NoStop}%
\bibitem [{\citenamefont {Campbell}\ \emph {et~al.}(2018)\citenamefont
  {Campbell}, \citenamefont {Ciccarello}, \citenamefont {Palma},\ and\
  \citenamefont {Vacchini}}]{CampbellPRA2018}%
  \BibitemOpen
  \bibfield  {author} {\bibinfo {author} {\bibfnamefont {S.}~\bibnamefont
  {Campbell}}, \bibinfo {author} {\bibfnamefont {F.}~\bibnamefont
  {Ciccarello}}, \bibinfo {author} {\bibfnamefont {G.~M.}\ \bibnamefont
  {Palma}}, \ and\ \bibinfo {author} {\bibfnamefont {B.}~\bibnamefont
  {Vacchini}},\ }\bibfield  {title} {\enquote {\bibinfo {title}
  {System-environment correlations and markovian embedding of quantum
  non-markovian dynamics},}\ }\href {\doibase 10.1103/PhysRevA.98.012142}
  {\bibfield  {journal} {\bibinfo  {journal} {Phys. Rev. A}\ }\textbf {\bibinfo
  {volume} {98}},\ \bibinfo {pages} {012142} (\bibinfo {year}
  {2018})}\BibitemShut {NoStop}%
\bibitem [{\citenamefont {Rau}(1963)}]{rau}%
  \BibitemOpen
  \bibfield  {author} {\bibinfo {author} {\bibfnamefont {J.}~\bibnamefont
  {Rau}},\ }\bibfield  {title} {\enquote {\bibinfo {title} {Relaxation
  phenomena in spin and harmonic oscillator systems},}\ }\href {\doibase
  10.1103/PhysRev.129.1880} {\bibfield  {journal} {\bibinfo  {journal} {Phys.
  Rev.}\ }\textbf {\bibinfo {volume} {129}},\ \bibinfo {pages} {1880--1888}
  (\bibinfo {year} {1963})}\BibitemShut {NoStop}%
\bibitem [{\citenamefont {Brun}(2002)}]{brun}%
  \BibitemOpen
  \bibfield  {author} {\bibinfo {author} {\bibfnamefont {T.~A.}\ \bibnamefont
  {Brun}},\ }\bibfield  {title} {\enquote {\bibinfo {title} {A simple model of
  quantum trajectories},}\ }\href {\doibase 10.1119/1.1475328} {\bibfield
  {journal} {\bibinfo  {journal} {Am. J. Phys.}\ }\textbf {\bibinfo {volume}
  {70}},\ \bibinfo {pages} {719--737} (\bibinfo {year} {2002})}\BibitemShut
  {NoStop}%
\bibitem [{\citenamefont {Scarani}\ \emph {et~al.}(2002)\citenamefont
  {Scarani}, \citenamefont {Ziman}, \citenamefont {\ifmmode \check{S}\else
  \v{S}\fi{}telmachovi\ifmmode~\check{c}\else \v{c}\fi{}}, \citenamefont
  {Gisin},\ and\ \citenamefont {Bu\ifmmode~\check{z}\else
  \v{z}\fi{}ek}}]{ScaraniPRL2002}%
  \BibitemOpen
  \bibfield  {author} {\bibinfo {author} {\bibfnamefont {V.}~\bibnamefont
  {Scarani}}, \bibinfo {author} {\bibfnamefont {M.}~\bibnamefont {Ziman}},
  \bibinfo {author} {\bibfnamefont {P.}~\bibnamefont {\ifmmode \check{S}\else
  \v{S}\fi{}telmachovi\ifmmode~\check{c}\else \v{c}\fi{}}}, \bibinfo {author}
  {\bibfnamefont {N.}~\bibnamefont {Gisin}}, \ and\ \bibinfo {author}
  {\bibfnamefont {V.}~\bibnamefont {Bu\ifmmode~\check{z}\else \v{z}\fi{}ek}},\
  }\bibfield  {title} {\enquote {\bibinfo {title} {Thermalizing quantum
  machines: Dissipation and entanglement},}\ }\href {\doibase
  10.1103/PhysRevLett.88.097905} {\bibfield  {journal} {\bibinfo  {journal}
  {Phys. Rev. Lett.}\ }\textbf {\bibinfo {volume} {88}},\ \bibinfo {pages}
  {097905} (\bibinfo {year} {2002})}\BibitemShut {NoStop}%
\bibitem [{\citenamefont {Ziman}\ and\ \citenamefont {Bu\ifmmode~\check{z}\else
  \v{z}\fi{}ek}(2005)}]{BuzekPRA2005}%
  \BibitemOpen
  \bibfield  {author} {\bibinfo {author} {\bibfnamefont {M.}~\bibnamefont
  {Ziman}}\ and\ \bibinfo {author} {\bibfnamefont {V.}~\bibnamefont
  {Bu\ifmmode~\check{z}\else \v{z}\fi{}ek}},\ }\bibfield  {title} {\enquote
  {\bibinfo {title} {All (qubit) decoherences: Complete characterization and
  physical implementation},}\ }\href {\doibase 10.1103/PhysRevA.72.022110}
  {\bibfield  {journal} {\bibinfo  {journal} {Phys. Rev. A}\ }\textbf {\bibinfo
  {volume} {72}},\ \bibinfo {pages} {022110} (\bibinfo {year}
  {2005})}\BibitemShut {NoStop}%
\bibitem [{\citenamefont {Hughes}\ \emph {et~al.}(2009)\citenamefont {Hughes},
  \citenamefont {Christ},\ and\ \citenamefont {Burghardt}}]{Hughes2009a}%
  \BibitemOpen
  \bibfield  {author} {\bibinfo {author} {\bibfnamefont {K.~H.}\ \bibnamefont
  {Hughes}}, \bibinfo {author} {\bibfnamefont {C.~D.}\ \bibnamefont {Christ}},
  \ and\ \bibinfo {author} {\bibfnamefont {I.}~\bibnamefont {Burghardt}},\
  }\bibfield  {title} {\enquote {\bibinfo {title} {Effective-mode
  representation of non-markovian dynamics: A hierarchical approximation of the
  spectral density. i. application to single surface dynamics},}\ }\href
  {\doibase 10.1063/1.3159671} {\bibfield  {journal} {\bibinfo  {journal} {J.
  Chem. Phys.}\ }\textbf {\bibinfo {volume} {131}},\ \bibinfo {pages} {024109}
  (\bibinfo {year} {2009})}\BibitemShut {NoStop}%
\bibitem [{\citenamefont {Chin}\ \emph {et~al.}(2010)\citenamefont {Chin},
  \citenamefont {Rivas}, \citenamefont {Huelga},\ and\ \citenamefont
  {Plenio}}]{Chin2010a}%
  \BibitemOpen
  \bibfield  {author} {\bibinfo {author} {\bibfnamefont {A.~W.}\ \bibnamefont
  {Chin}}, \bibinfo {author} {\bibfnamefont {A.}~\bibnamefont {Rivas}},
  \bibinfo {author} {\bibfnamefont {S.~F.}\ \bibnamefont {Huelga}}, \ and\
  \bibinfo {author} {\bibfnamefont {M.~B.}\ \bibnamefont {Plenio}},\ }\bibfield
   {title} {\enquote {\bibinfo {title} {Exact mapping between system-reservoir
  quantum models and semi-infinite discrete chains using orthogonal
  polynomials},}\ }\href {\doibase http://dx.doi.org/10.1063/1.3490188}
  {\bibfield  {journal} {\bibinfo  {journal} {J. Math. Phys.}\ }\textbf
  {\bibinfo {volume} {51}},\ \bibinfo {eid} {092109} (\bibinfo {year}
  {2010})}\BibitemShut {NoStop}%
\bibitem [{\citenamefont {Martinazzo}\ \emph {et~al.}(2011)\citenamefont
  {Martinazzo}, \citenamefont {Vacchini}, \citenamefont {Hughes},\ and\
  \citenamefont {Burghardt}}]{Martinazzo2011a}%
  \BibitemOpen
  \bibfield  {author} {\bibinfo {author} {\bibfnamefont {R.}~\bibnamefont
  {Martinazzo}}, \bibinfo {author} {\bibfnamefont {B.}~\bibnamefont
  {Vacchini}}, \bibinfo {author} {\bibfnamefont {K.~H.}\ \bibnamefont
  {Hughes}}, \ and\ \bibinfo {author} {\bibfnamefont {I.}~\bibnamefont
  {Burghardt}},\ }\bibfield  {title} {\enquote {\bibinfo {title} {Universal
  markovian reduction of brownian particle dynamics},}\ }\href {\doibase
  10.1063/1.3532408} {\bibfield  {journal} {\bibinfo  {journal} {J. Chem.
  Phys.}\ }\textbf {\bibinfo {volume} {134}},\ \bibinfo {pages} {011101}
  (\bibinfo {year} {2011})}\BibitemShut {NoStop}%
\bibitem [{\citenamefont {Woods}\ \emph {et~al.}(2014)\citenamefont {Woods},
  \citenamefont {Groux}, \citenamefont {Chin}, \citenamefont {Huelga},\ and\
  \citenamefont {Plenio}}]{Woods2014a}%
  \BibitemOpen
  \bibfield  {author} {\bibinfo {author} {\bibfnamefont {M.~P.}\ \bibnamefont
  {Woods}}, \bibinfo {author} {\bibfnamefont {R.}~\bibnamefont {Groux}},
  \bibinfo {author} {\bibfnamefont {A.~W.}\ \bibnamefont {Chin}}, \bibinfo
  {author} {\bibfnamefont {S.~F.}\ \bibnamefont {Huelga}}, \ and\ \bibinfo
  {author} {\bibfnamefont {M.~B.}\ \bibnamefont {Plenio}},\ }\bibfield  {title}
  {\enquote {\bibinfo {title} {Mappings of open quantum systems onto chain
  representations and markovian embeddings},}\ }\href {\doibase
  10.1063/1.4866769} {\bibfield  {journal} {\bibinfo  {journal} {J. Math.
  Phys.}\ }\textbf {\bibinfo {volume} {55}},\ \bibinfo {pages} {032101}
  (\bibinfo {year} {2014})}\BibitemShut {NoStop}%
\bibitem [{\citenamefont {Strasberg}\ \emph {et~al.}(2016)\citenamefont
  {Strasberg}, \citenamefont {Schaller}, \citenamefont {Lambert},\ and\
  \citenamefont {Brandes}}]{Strasberg2016a}%
  \BibitemOpen
  \bibfield  {author} {\bibinfo {author} {\bibfnamefont {P.}~\bibnamefont
  {Strasberg}}, \bibinfo {author} {\bibfnamefont {G.}~\bibnamefont {Schaller}},
  \bibinfo {author} {\bibfnamefont {N.}~\bibnamefont {Lambert}}, \ and\
  \bibinfo {author} {\bibfnamefont {T.}~\bibnamefont {Brandes}},\ }\bibfield
  {title} {\enquote {\bibinfo {title} {Nonequilibrium thermodynamics in the
  strong coupling and non-markovian regime based on a reaction coordinate
  mapping},}\ }\href {http://stacks.iop.org/1367-2630/18/i=7/a=073007}
  {\bibfield  {journal} {\bibinfo  {journal} {New J. Phys.}\ }\textbf {\bibinfo
  {volume} {18}},\ \bibinfo {pages} {073007} (\bibinfo {year}
  {2016})}\BibitemShut {NoStop}%
\bibitem [{\citenamefont {Tamascelli}\ \emph
  {et~al.}(2018{\natexlab{a}})\citenamefont {Tamascelli}, \citenamefont
  {Smirne}, \citenamefont {Huelga},\ and\ \citenamefont {Plenio}}]{Tama1}%
  \BibitemOpen
  \bibfield  {author} {\bibinfo {author} {\bibfnamefont {D.}~\bibnamefont
  {Tamascelli}}, \bibinfo {author} {\bibfnamefont {A.}~\bibnamefont {Smirne}},
  \bibinfo {author} {\bibfnamefont {S.~F.}\ \bibnamefont {Huelga}}, \ and\
  \bibinfo {author} {\bibfnamefont {M.~B.}\ \bibnamefont {Plenio}},\ }\bibfield
   {title} {\enquote {\bibinfo {title} {Nonperturbative treatment of
  non-markovian dynamics of open quantum systems},}\ }\href {\doibase
  10.1103/PhysRevLett.120.030402} {\bibfield  {journal} {\bibinfo  {journal}
  {Phys. Rev. Lett.}\ }\textbf {\bibinfo {volume} {120}},\ \bibinfo {pages}
  {030402} (\bibinfo {year} {2018}{\natexlab{a}})}\BibitemShut {NoStop}%
\bibitem [{\citenamefont {Tamascelli}\ \emph
  {et~al.}(2018{\natexlab{b}})\citenamefont {Tamascelli}, \citenamefont
  {Smirne}, \citenamefont {Huelga},\ and\ \citenamefont {Plenio}}]{Tama2}%
  \BibitemOpen
  \bibfield  {author} {\bibinfo {author} {\bibfnamefont {D.}~\bibnamefont
  {Tamascelli}}, \bibinfo {author} {\bibfnamefont {A.}~\bibnamefont {Smirne}},
  \bibinfo {author} {\bibfnamefont {S.~F.}\ \bibnamefont {Huelga}}, \ and\
  \bibinfo {author} {\bibfnamefont {M.~B.}\ \bibnamefont {Plenio}},\ }\bibfield
   {title} {\enquote {\bibinfo {title} {Efficient simulation of
  finite-temperature open quantum systems},}\ }\href
  {https://arxiv.org/abs/1811.12418} {\bibfield  {journal} {\bibinfo  {journal}
  {arXiv:1811.12418}\ } (\bibinfo {year} {2018}{\natexlab{b}})}\BibitemShut
  {NoStop}%
\bibitem [{\citenamefont {Fuchs}\ and\ \citenamefont {{van de
  Graaf}}(1999)}]{Fuchs1999a}%
  \BibitemOpen
  \bibfield  {author} {\bibinfo {author} {\bibfnamefont {C.~A.}\ \bibnamefont
  {Fuchs}}\ and\ \bibinfo {author} {\bibfnamefont {J.}~\bibnamefont {{van de
  Graaf}}},\ }\bibfield  {title} {\enquote {\bibinfo {title} {Cryptographic
  distinguishability measures for quantum-mechanical states},}\ }\href
  {\doibase 10.1109/18.761271} {\bibfield  {journal} {\bibinfo  {journal} {IEEE
  Trans. Inf. Th.}\ }\textbf {\bibinfo {volume} {45}},\ \bibinfo {pages}
  {1216--1227} (\bibinfo {year} {1999})}\BibitemShut {NoStop}%
\bibitem [{\citenamefont {Vasile}\ \emph {et~al.}(2011)\citenamefont {Vasile},
  \citenamefont {Maniscalco}, \citenamefont {Paris}, \citenamefont {Breuer},\
  and\ \citenamefont {Piilo}}]{Vasile2011a}%
  \BibitemOpen
  \bibfield  {author} {\bibinfo {author} {\bibfnamefont {R.}~\bibnamefont
  {Vasile}}, \bibinfo {author} {\bibfnamefont {S.}~\bibnamefont {Maniscalco}},
  \bibinfo {author} {\bibfnamefont {M.~G.~A.}\ \bibnamefont {Paris}}, \bibinfo
  {author} {\bibfnamefont {H.-P.}\ \bibnamefont {Breuer}}, \ and\ \bibinfo
  {author} {\bibfnamefont {J.}~\bibnamefont {Piilo}},\ }\bibfield  {title}
  {\enquote {\bibinfo {title} {Quantifying non-markovianity of
  continuous-variable gaussian dynamical maps},}\ }\href {\doibase
  10.1103/PhysRevA.84.052118} {\bibfield  {journal} {\bibinfo  {journal} {Phys.
  Rev. A}\ }\textbf {\bibinfo {volume} {84}},\ \bibinfo {pages} {052118}
  (\bibinfo {year} {2011})}\BibitemShut {NoStop}%
\bibitem [{\citenamefont {Olivares}(2012)}]{OlivaresEPJST2012}%
  \BibitemOpen
  \bibfield  {author} {\bibinfo {author} {\bibfnamefont {S.}~\bibnamefont
  {Olivares}},\ }\bibfield  {title} {\enquote {\bibinfo {title} {Quantum optics
  in the phase space},}\ }\href {\doibase 10.1140/epjst/e2012-01532-4}
  {\bibfield  {journal} {\bibinfo  {journal} {Eur. Phys. J. Special Topics}\
  }\textbf {\bibinfo {volume} {203}},\ \bibinfo {pages} {3?24} (\bibinfo {year}
  {2012})}\BibitemShut {NoStop}%
\bibitem [{\citenamefont {Banchi}\ \emph {et~al.}(2015)\citenamefont {Banchi},
  \citenamefont {Braunstein},\ and\ \citenamefont {Pirandola}}]{BanchiPRL}%
  \BibitemOpen
  \bibfield  {author} {\bibinfo {author} {\bibfnamefont {L.}~\bibnamefont
  {Banchi}}, \bibinfo {author} {\bibfnamefont {S.~L.}\ \bibnamefont
  {Braunstein}}, \ and\ \bibinfo {author} {\bibfnamefont {S.}~\bibnamefont
  {Pirandola}},\ }\bibfield  {title} {\enquote {\bibinfo {title} {Quantum
  fidelity for arbitrary gaussian states},}\ }\href {\doibase
  10.1103/PhysRevLett.115.260501} {\bibfield  {journal} {\bibinfo  {journal}
  {Phys. Rev. Lett.}\ }\textbf {\bibinfo {volume} {115}},\ \bibinfo {pages}
  {260501} (\bibinfo {year} {2015})}\BibitemShut {NoStop}%
\bibitem [{\citenamefont {Wang}\ and\ \citenamefont
  {Schirmer}(2009)}]{WangPRA2009}%
  \BibitemOpen
  \bibfield  {author} {\bibinfo {author} {\bibfnamefont {X.}~\bibnamefont
  {Wang}}\ and\ \bibinfo {author} {\bibfnamefont {S.~G.}\ \bibnamefont
  {Schirmer}},\ }\bibfield  {title} {\enquote {\bibinfo {title} {Contractivity
  of the hilbert-schmidt distance under open-system dynamics},}\ }\href
  {\doibase 10.1103/PhysRevA.79.052326} {\bibfield  {journal} {\bibinfo
  {journal} {Phys. Rev. A}\ }\textbf {\bibinfo {volume} {79}},\ \bibinfo
  {pages} {052326} (\bibinfo {year} {2009})}\BibitemShut {NoStop}%
\bibitem [{\citenamefont {McCloskey}\ and\ \citenamefont
  {Paternostro}(2014)}]{McCloskeyPRA2014}%
  \BibitemOpen
  \bibfield  {author} {\bibinfo {author} {\bibfnamefont {R.}~\bibnamefont
  {McCloskey}}\ and\ \bibinfo {author} {\bibfnamefont {M.}~\bibnamefont
  {Paternostro}},\ }\bibfield  {title} {\enquote {\bibinfo {title}
  {Non-markovianity and system-environment correlations in a microscopic
  collision model},}\ }\href {\doibase 10.1103/PhysRevA.89.052120} {\bibfield
  {journal} {\bibinfo  {journal} {Phys. Rev. A}\ }\textbf {\bibinfo {volume}
  {89}},\ \bibinfo {pages} {052120} (\bibinfo {year} {2014})}\BibitemShut
  {NoStop}%
\bibitem [{\citenamefont {Kretschmer}\ \emph {et~al.}(2016)\citenamefont
  {Kretschmer}, \citenamefont {Luoma},\ and\ \citenamefont
  {Strunz}}]{StrunzPRA2016}%
  \BibitemOpen
  \bibfield  {author} {\bibinfo {author} {\bibfnamefont {S.}~\bibnamefont
  {Kretschmer}}, \bibinfo {author} {\bibfnamefont {K.}~\bibnamefont {Luoma}}, \
  and\ \bibinfo {author} {\bibfnamefont {W.~T.}\ \bibnamefont {Strunz}},\
  }\bibfield  {title} {\enquote {\bibinfo {title} {Collision model for
  non-markovian quantum dynamics},}\ }\href {\doibase
  10.1103/PhysRevA.94.012106} {\bibfield  {journal} {\bibinfo  {journal} {Phys.
  Rev. A}\ }\textbf {\bibinfo {volume} {94}},\ \bibinfo {pages} {012106}
  (\bibinfo {year} {2016})}\BibitemShut {NoStop}%
\bibitem [{\citenamefont {\ifmmode~\mbox{\c{C}}\else \c{C}\fi{}akmak}\ \emph
  {et~al.}(2017)\citenamefont {\ifmmode~\mbox{\c{C}}\else \c{C}\fi{}akmak},
  \citenamefont {Pezzutto}, \citenamefont {Paternostro},\ and\ \citenamefont
  {M\"ustecapl\ifmmode \imath \else \i \fi{}o\ifmmode~\breve{g}\else
  \u{g}\fi{}lu}}]{CakmakPRA2017b}%
  \BibitemOpen
  \bibfield  {author} {\bibinfo {author} {\bibfnamefont {B.}~\bibnamefont
  {\ifmmode~\mbox{\c{C}}\else \c{C}\fi{}akmak}}, \bibinfo {author}
  {\bibfnamefont {M.}~\bibnamefont {Pezzutto}}, \bibinfo {author}
  {\bibfnamefont {M.}~\bibnamefont {Paternostro}}, \ and\ \bibinfo {author}
  {\bibfnamefont {\"O.~E.}\ \bibnamefont {M\"ustecapl\ifmmode \imath \else \i
  \fi{}o\ifmmode~\breve{g}\else \u{g}\fi{}lu}},\ }\bibfield  {title} {\enquote
  {\bibinfo {title} {Non-markovianity, coherence, and system-environment
  correlations in a long-range collision model},}\ }\href {\doibase
  10.1103/PhysRevA.96.022109} {\bibfield  {journal} {\bibinfo  {journal} {Phys.
  Rev. A}\ }\textbf {\bibinfo {volume} {96}},\ \bibinfo {pages} {022109}
  (\bibinfo {year} {2017})}\BibitemShut {NoStop}%
\bibitem [{\citenamefont {Jin}\ and\ \citenamefont {Yu}(2018)}]{JinNJP2018}%
  \BibitemOpen
  \bibfield  {author} {\bibinfo {author} {\bibfnamefont {J.}~\bibnamefont
  {Jin}}\ and\ \bibinfo {author} {\bibfnamefont {C.-S.}\ \bibnamefont {Yu}},\
  }\bibfield  {title} {\enquote {\bibinfo {title} {Non-markovianity in a
  collision model with environmental block},}\ }\href {\doibase
  10.1088/1367-2630/aac0cb} {\bibfield  {journal} {\bibinfo  {journal} {New J.
  Phys.}\ }\textbf {\bibinfo {volume} {20}},\ \bibinfo {pages} {053026}
  (\bibinfo {year} {2018})}\BibitemShut {NoStop}%
\bibitem [{\citenamefont {Strasberg}\ \emph {et~al.}(2017)\citenamefont
  {Strasberg}, \citenamefont {Schaller}, \citenamefont {Brandes},\ and\
  \citenamefont {Esposito}}]{EspositoPRX}%
  \BibitemOpen
  \bibfield  {author} {\bibinfo {author} {\bibfnamefont {P.}~\bibnamefont
  {Strasberg}}, \bibinfo {author} {\bibfnamefont {G.}~\bibnamefont {Schaller}},
  \bibinfo {author} {\bibfnamefont {T.}~\bibnamefont {Brandes}}, \ and\
  \bibinfo {author} {\bibfnamefont {M.}~\bibnamefont {Esposito}},\ }\bibfield
  {title} {\enquote {\bibinfo {title} {Quantum and information thermodynamics:
  A unifying framework based on repeated interactions},}\ }\href {\doibase
  10.1103/PhysRevX.7.021003} {\bibfield  {journal} {\bibinfo  {journal} {Phys.
  Rev. X}\ }\textbf {\bibinfo {volume} {7}},\ \bibinfo {pages} {021003}
  (\bibinfo {year} {2017})}\BibitemShut {NoStop}%
\bibitem [{\citenamefont {Ciccarello}(2018)}]{FrancescoQMQM}%
  \BibitemOpen
  \bibfield  {author} {\bibinfo {author} {\bibfnamefont {F.}~\bibnamefont
  {Ciccarello}},\ }\bibfield  {title} {\enquote {\bibinfo {title} {Collision
  models in quantum optics},}\ }\href {\doibase 10.1515/qmetro-2017-0007}
  {\bibfield  {journal} {\bibinfo  {journal} {Quantum Meas. Quantum Metrol.}\
  }\textbf {\bibinfo {volume} {4}},\ \bibinfo {pages} {53} (\bibinfo {year}
  {2018})}\BibitemShut {NoStop}%
\bibitem [{\citenamefont {{Pezzutto}}\ \emph {et~al.}(2016)\citenamefont
  {{Pezzutto}}, \citenamefont {{Paternostro}},\ and\ \citenamefont
  {{Omar}}}]{PezzuttoNJP}%
  \BibitemOpen
  \bibfield  {author} {\bibinfo {author} {\bibfnamefont {M.}~\bibnamefont
  {{Pezzutto}}}, \bibinfo {author} {\bibfnamefont {M.}~\bibnamefont
  {{Paternostro}}}, \ and\ \bibinfo {author} {\bibfnamefont {Y.}~\bibnamefont
  {{Omar}}},\ }\bibfield  {title} {\enquote {\bibinfo {title} {{Implications of
  non-Markovian quantum dynamics for the Landauer bound}},}\ }\href {\doibase
  10.1088/1367-2630/18/12/123018} {\bibfield  {journal} {\bibinfo  {journal}
  {New J. Phys.}\ }\textbf {\bibinfo {volume} {18}},\ \bibinfo {pages} {123018}
  (\bibinfo {year} {2016})}\BibitemShut {NoStop}%
\bibitem [{\citenamefont {Chiuri}\ \emph {et~al.}(2012)\citenamefont {Chiuri},
  \citenamefont {Greganti}, \citenamefont {Mazzola}, \citenamefont
  {Paternostro},\ and\ \citenamefont {Mataloni}}]{SciRepExp}%
  \BibitemOpen
  \bibfield  {author} {\bibinfo {author} {\bibfnamefont {A.}~\bibnamefont
  {Chiuri}}, \bibinfo {author} {\bibfnamefont {C.}~\bibnamefont {Greganti}},
  \bibinfo {author} {\bibfnamefont {L.}~\bibnamefont {Mazzola}}, \bibinfo
  {author} {\bibfnamefont {M.}~\bibnamefont {Paternostro}}, \ and\ \bibinfo
  {author} {\bibfnamefont {P.}~\bibnamefont {Mataloni}},\ }\bibfield  {title}
  {\enquote {\bibinfo {title} {Linear optics simulation of quantum
  non-markovian dynamics},}\ }\href {\doibase 10.1038/srep00968} {\bibfield
  {journal} {\bibinfo  {journal} {Sci. Rep.}\ }\textbf {\bibinfo {volume}
  {2}},\ \bibinfo {pages} {968} (\bibinfo {year} {2012})}\BibitemShut {NoStop}%
\bibitem [{\citenamefont {Ciampini}\ \emph {et~al.}(2018)\citenamefont
  {Ciampini}, \citenamefont {Pinna}, \citenamefont {Mataloni},\ and\
  \citenamefont {Paternostro}}]{DarwinExp}%
  \BibitemOpen
  \bibfield  {author} {\bibinfo {author} {\bibfnamefont {M.~A.}\ \bibnamefont
  {Ciampini}}, \bibinfo {author} {\bibfnamefont {G.}~\bibnamefont {Pinna}},
  \bibinfo {author} {\bibfnamefont {P.}~\bibnamefont {Mataloni}}, \ and\
  \bibinfo {author} {\bibfnamefont {M.}~\bibnamefont {Paternostro}},\
  }\bibfield  {title} {\enquote {\bibinfo {title} {Experimental signature of
  quantum darwinism in photonic cluster states},}\ }\href {\doibase
  10.1103/PhysRevA.98.020101} {\bibfield  {journal} {\bibinfo  {journal} {Phys.
  Rev. A}\ }\textbf {\bibinfo {volume} {98}},\ \bibinfo {pages} {020101}
  (\bibinfo {year} {2018})}\BibitemShut {NoStop}%
\bibitem [{\citenamefont {Cuevas}\ \emph {et~al.}(2019)\citenamefont {Cuevas},
  \citenamefont {Geraldi}, \citenamefont {Liorni}, \citenamefont {Bonavena},
  \citenamefont {Pasquale}, \citenamefont {Sciarrino}, \citenamefont
  {Giovannetti},\ and\ \citenamefont {Mataloni}}]{MataloniExp}%
  \BibitemOpen
  \bibfield  {author} {\bibinfo {author} {\bibfnamefont {A.}~\bibnamefont
  {Cuevas}}, \bibinfo {author} {\bibfnamefont {A.}~\bibnamefont {Geraldi}},
  \bibinfo {author} {\bibfnamefont {C.}~\bibnamefont {Liorni}}, \bibinfo
  {author} {\bibfnamefont {L.~D.}\ \bibnamefont {Bonavena}}, \bibinfo {author}
  {\bibfnamefont {A.~De}\ \bibnamefont {Pasquale}}, \bibinfo {author}
  {\bibfnamefont {F.}~\bibnamefont {Sciarrino}}, \bibinfo {author}
  {\bibfnamefont {V.}~\bibnamefont {Giovannetti}}, \ and\ \bibinfo {author}
  {\bibfnamefont {P.}~\bibnamefont {Mataloni}},\ }\bibfield  {title} {\enquote
  {\bibinfo {title} {All-optical implementation of collision-based evolutions
  of open quantum systems},}\ }\href {\doibase 10.1038/s41598-019-39832-9}
  {\bibfield  {journal} {\bibinfo  {journal} {Sci. Rep.}\ }\textbf {\bibinfo
  {volume} {9}},\ \bibinfo {pages} {3205} (\bibinfo {year} {2019})}\BibitemShut
  {NoStop}%
\bibitem [{\citenamefont {Cusumano}\ \emph {et~al.}(2018)\citenamefont
  {Cusumano}, \citenamefont {Cavina}, \citenamefont {Keck}, \citenamefont
  {De~Pasquale},\ and\ \citenamefont {Giovannetti}}]{GiovannettiPRA2018}%
  \BibitemOpen
  \bibfield  {author} {\bibinfo {author} {\bibfnamefont {S.}~\bibnamefont
  {Cusumano}}, \bibinfo {author} {\bibfnamefont {V.}~\bibnamefont {Cavina}},
  \bibinfo {author} {\bibfnamefont {M.}~\bibnamefont {Keck}}, \bibinfo {author}
  {\bibfnamefont {A.}~\bibnamefont {De~Pasquale}}, \ and\ \bibinfo {author}
  {\bibfnamefont {V.}~\bibnamefont {Giovannetti}},\ }\bibfield  {title}
  {\enquote {\bibinfo {title} {Entropy production and asymptotic factorization
  via thermalization: A collisional model approach},}\ }\href {\doibase
  10.1103/PhysRevA.98.032119} {\bibfield  {journal} {\bibinfo  {journal} {Phys.
  Rev. A}\ }\textbf {\bibinfo {volume} {98}},\ \bibinfo {pages} {032119}
  (\bibinfo {year} {2018})}\BibitemShut {NoStop}%
\end{thebibliography}%

\end{document}